\documentclass[article]{IEEEtran}

\usepackage{graphicx,colordvi,psfrag}
\usepackage{amsmath,amssymb}
\usepackage{epstopdf}
\usepackage[caption=false]{subfig}
\usepackage{epsfig,cite}
\usepackage{calc,pstricks, pgf, xcolor}
\usepackage{nicefrac}
\usepackage{enumerate}
\usepackage{bm}
\usepackage{fixltx2e}
\usepackage{pstricks-add}

\newtheorem{theorem}{Theorem}

\newtheorem{remark}{Remark}
\newtheorem{example}{Example}
\newtheorem{definition}{Definition}
\newtheorem{lemma}{Lemma}
\newtheorem{proposition}{Proposition}
\newenvironment{proof}[1][Proof]{\noindent\textbf{#1.} }{\ \rule{0.5em}{0.5em}}

\newcommand{\by}{\mathbf{y}}
\newcommand{\bY}{\mathbf{Y}}
\newcommand{\bx}{\mathbf{x}}
\newcommand{\bX}{\mathbf{X}}
\newcommand{\bz}{\mathbf{z}}
\newcommand{\bZ}{\mathbf{Z}}

\newcommand{\bH}{\mathbf{H}}
\newcommand{\bA}{\mathbf{A}}
\newcommand{\ba}{\mathbf{a}}

\newcommand{\bd}{\mathbf{d}}
\newcommand{\bD}{\mathbf{D}}
\newcommand{\bB}{\mathbf{B}}
\newcommand{\bb}{\mathbf{b}}

\newcommand{\bu}{\mathbf{u}}

\newcommand{\bv}{\mathbf{v}}
\newcommand{\CV}{\mathcal{V}}

\newcommand{\be}{\mathbf{e}}

\newcommand{\bI}{\mathbf{I}}

\newcommand{\bG}{\mathbf{G}}

\newcommand{\bK}{\mathbf{K}}
\newcommand{\bs}{\mathbf{s}}
\newcommand{\bS}{\mathbf{S}}

\newcommand{\bF}{\mathbf{F}}

\newcommand{\ZZ}{\mathbb{Z}}

\newcommand{\RR}{\mathbb{R}}

\newcommand{\Tsnr}{\mathsf{SNR}}

\newcommand{\Ql}{Q_{\Lambda}}
\newcommand{\Qlf}{Q_{\Lambda_f}}
\newcommand{\Mod}{\bmod\Lambda}

\newcommand{\Span}{\mathop{\mathrm{span}}}

\newcommand{\Unif}{\mathop{\mathrm{Unif}}}
\newcommand{\Vol}{\mathrm{Vol}}

\DeclareMathOperator*{\argmin}{\arg\!\min}

\newrgbcolor{BoxRed}{1 .5 .5}
\newrgbcolor{BoxBlue}{.9 .9 1}
\newrgbcolor{lightgrey}{.7 .7 0.7}

\begin{document}

\title{Integer-Forcing Source Coding}
\author{Or~Ordentlich and
        Uri~Erez,~\IEEEmembership{Member,~IEEE}
\thanks{The work of U. Erez was supported in part by the Israel Science Foundation under Grant No. 1557/13. The work of O. Ordentlich was supported by the Adams Fellowship Program of the Israel Academy of Sciences and Humanities, and a fellowship from The Yitzhak and Chaya Weinstein Research Institute for Signal Processing at Tel Aviv University.}
\thanks{O. Ordentlich and U. Erez are with Tel Aviv University, Tel Aviv, Israel (email: ordent,uri@eng.tau.ac.il).
}}


\maketitle

\begin{abstract}
Integer-Forcing (IF) is a new framework, based on compute-and-forward, for decoding multiple integer linear combinations from the output of a Gaussian multiple-input multiple-output channel. 
This work applies the IF approach to arrive at a new low-complexity scheme, IF source coding, for distributed lossy compression of correlated Gaussian sources under a minimum mean squared error distortion measure. All encoders use the same nested lattice codebook. Each encoder quantizes its observation using the fine lattice as a quantizer and reduces the result modulo the coarse lattice, which plays the role of binning. Rather than directly recovering the individual quantized signals, the decoder first recovers a full-rank set of judiciously chosen integer linear combinations of the quantized signals, and then inverts it. In general, the linear combinations have smaller average powers than the original signals. This allows to increase the density of the coarse lattice, which in turn translates to smaller compression rates. We also propose and analyze a one-shot version of IF source coding, that is simple enough to potentially lead to a new design principle for analog-to-digital converters that can exploit spatial correlations between the sampled signals.
\end{abstract}

\section{Introduction}
\label{sec:intro}

\begin{figure}[]
\begin{center}
\psset{unit=0.6mm}
\begin{pspicture}(0,0)(92,70)

\rput(0,50){
\rput(0,0){$\bx_1$} \psline{->}(5,0)(12,0)
\psframe(12,-4)(28,4)
\rput(21,0){$\mathcal{E}_1$}
\psline{->}(28,0)(50,0)
\psline(38.5,2)(42.5,-2)
\rput(40,5){$R_1$}

}

\rput(0,33){\rput(3,2){$\vdots$}}

\rput(0,10){\rput(0,0){$\bx_K$} \psline{->}(5,0)(12,0)
\psframe(12,-4)(28,4)
\rput(21,0){$\mathcal{E}_K$}
\psline{->}(28,0)(50,0)
\psline(38.5,2)(42.5,-2)
\rput(40,5){$R_K$}

}

\rput(50,0){
\psframe(0,0)(15,60)\rput(7.5,30){$\mathcal{D}$}
\psline{->}(15,30)(22,30)
\rput(35,30){$\begin{array}{c}
               (\hat{\bx}_1,d_1) \\
               \vdots \\
               (\hat{\bx}_K,d_K)
             \end{array}$
}
}

\end{pspicture}
\end{center}
\caption{The distributed source coding problem. The $k$th encoder $\mathcal{E}_k$ has access to the vector $\bx_k$ that contains $n$ i.i.d. realizations of the random variable $x_k$. It encodes $\bx_k$ to an index taking values in $1,\ldots,2^{n R_k}$. The sources $x_1,\ldots,x_K$ are assumed correlated and the encoders are not allowed to cooperate. The decoder's goal is to produce estimates of each $\bx_k$ with average distortions $d_k$ using the $K$ indices it received from the encoders.}
\label{fig:distributedSC}
\end{figure}
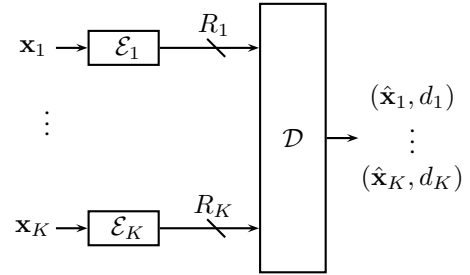

The distributed lossy compression problem, depicted in Figure~\ref{fig:distributedSC}, consists of multiple distributed encoders and one decoder. The encoders have access to correlated observations which they try to describe to the decoder with minimum rate and minimum distortion~\cite{tungphd,berger77,ElGamalKim}. This problem naturally arises in numerous scenarios.
For instance, consider a sensor network where multiple sensors that observe correlated random variables are connected via finite rate links to a central processor, but not to one another, and have to describe their observations to the central processor with minimum distortion. As another example, consider two competing television channels that cover the same event and have to broadcast their programs to the same end-users (that may choose which channel to watch and therefore need to be able to recover both programs with low distortion). Although the distributed lossy compression problem is usually classified as a pure source-coding problem, it is also an important building block in network channel coding problems. For instance, multiple relays may observe correlated signals that describe the messages transmitted by the different encoders in the network. The relays can compress-and-forward these signals further down the network in order to ultimately help the decoder recover the transmitted messages.

A special case that received considerable attention is that of distributed lossy compression of jointly Gaussian random variables under a quadratic distortion measure. The best known achievable scheme is that of Berger and Tung~\cite{tungphd,berger77}, although some examples where Berger-Tung compression can be outperformed are known~\cite{ng07IT,kp09,wagner11}. In the Gaussian case, the Berger-Tung approach reduces to each encoder compressing its source using a standard point-to-point quantizer, followed by Slepian-Wolf~\cite{sw73} encoding. For the quadratic Gaussian case with $K=2$, Wagner \textit{et al}.~\cite{wtv08} proved that this approach is optimal.

The importance of the quadratic-Gaussian distributed lossy compression problem has motivated researchers to design low-complexity encoding schemes that approach the performance of the Berger-Tung inner bound. This line of work was pioneered in~\cite{pr03,pr00} and remains an active area of research, see, e.g.,~\cite{xlc04,kw07,zse02} and references therein. However, at a high level, the existing approaches for distributed source coding are either notably asymmetric in the rates they require from the encoders, as they rely on the lattice-based implementation of Wyner-Ziv coding~\cite{zs98,zse02} and successive Wyner-Ziv coding~\cite{xlc04}, or specifically tailored to predefined correlation characteristics of the sources~\cite{pr00}. In general, the rate requirements in schemes that are based on Wyner-Ziv coding can be symmetrized by time-sharing between different compression/decompression orders~\cite{zse02}. Nevertheless, schemes using time-sharing have a few drawbacks. First, it requires the encoders and the decoders to use a larger number of codebooks, which complicates implementation. Second, it requires coordination between the distributed encoders, which is less crucial when time-sharing is not used. Finally, the compression block must be at least as long as the number of operation points that are time-shared.

In this work we propose a novel framework, \emph{integer-forcing source coding}, for distributed lossy compression with \emph{symmetric} rate and distortion requirements for all encoders. This scheme does not incorporate time-sharing. As in previous works, our approach is based on standard quantization followed by lattice-based binning. However, in contrast to previous works, in the proposed framework the decoder first uses the bin indices for recovering linear combinations with integer coefficients of the quantized signals, and only then recovers the quantized signals themselves. The decoder is free to optimize the full-rank set of integer-valued coefficients such as to best exploit the correlations between the quantized signals. Choosing these coefficients appropriately results in performance that is close to that of a joint typicality decoder, with a substantially smaller computational burden. In fact, the only operations performed by the encoders are quantization and lattice-binning which corresponds to nearest neighbor decoding, whereas the decoder is only required to perform matrix multiplications and nearest neighbor decoding operations.

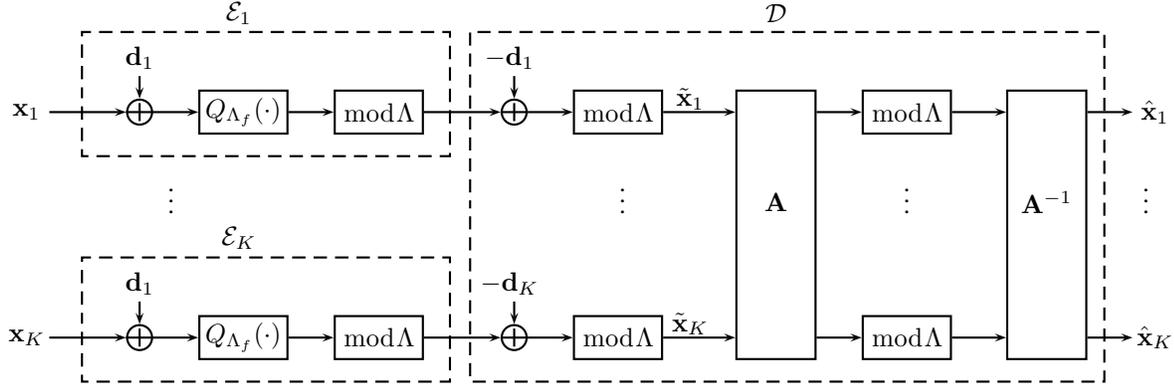
\begin{figure*}[]
\begin{center}
\psset{unit=0.6mm}
\begin{pspicture}(0,-12)(280,70)

\rput(0,50){
\rput(3,0){$\bx_1$} \psline{->}(8,0)(25,0)
\rput(25,0){
\pscircle(3,0){3}
\psline(0,0)(6,0)
\psline(3,-2)(3,2)
\psline{->}(3,8)(3,3)
\rput(3,12){$\bd_1$}
}
\rput(31,0){
\psline{->}(0,0)(10,0)
\psframe(10,-5)(30,5)\rput(20,0){$Q_{\Lambda_f}(\cdot)$}
}
\rput(61,0){\psline{->}(0,0)(10,0)
\psframe(10,-5)(30,5)\rput(20,0){$\Mod$}
}
\psline{->}(91,0)(108,0)

\rput(50,22){${\mathcal{E}_1}$}
\psframe[linestyle=dashed](15,-10)(97,18)

\rput(108,0){
\pscircle(3,0){3}
\psline(0,0)(6,0)
\psline(3,-2)(3,2)
\psline{->}(3,8)(3,3)
\rput(2,12){$-\bd_1$}
}
\rput(114,0){\psline{->}(0,0)(10,0)
\psframe(10,-5)(30,5)\rput(20,0){$\Mod$}
\psline{->}(30,0)(46,0)
\rput(36,3){$\tilde{\bx}_1$}
}

}

\rput(0,30){\rput(35,2){$\vdots$}}

\rput(100,30){\rput(35,2){$\vdots$}}

\rput(0,0){
\rput(3,0){$\bx_K$} \psline{->}(8,0)(25,0)
\rput(25,0){
\pscircle(3,0){3}
\psline(0,0)(6,0)
\psline(3,-2)(3,2)
\psline{->}(3,8)(3,3)
\rput(3,12){$\bd_1$}
}
\rput(31,0){
\psline{->}(0,0)(10,0)
\psframe(10,-5)(30,5)\rput(20,0){$Q_{\Lambda_f}(\cdot)$}
}
\rput(61,0){\psline{->}(0,0)(10,0)
\psframe(10,-5)(30,5)\rput(20,0){$\Mod$}
}
\psline{->}(91,0)(108,0)

\rput(50,22){${\mathcal{E}_K}$}
\psframe[linestyle=dashed](15,-10)(97,18)

\rput(108,0){
\pscircle(3,0){3}
\psline(0,0)(6,0)
\psline(3,-2)(3,2)
\psline{->}(3,8)(3,3)
\rput(2,12){$-\bd_K$}
}
\rput(114,0){\psline{->}(0,0)(10,0)
\psframe(10,-5)(30,5)\rput(20,0){$\Mod$}
\psline{->}(30,0)(46,0)
\rput(36,3){$\tilde{\bx}_K$}
}
}

\rput(160,0){
\psframe(0,-5)(18,55)\rput(9,30){$\bA$}}

\rput(178,50){
\psline{->}(0,0)(10,0)
\psframe(10,-5)(30,5)\rput(20,0){$\Mod$}
\psline{->}(30,0)(42,0)
}

\rput(178,30){\rput(20,2){$\vdots$}}

\rput(178,0){
\psline{->}(0,0)(10,0)
\psframe(10,-5)(30,5)\rput(20,0){$\Mod$}
\psline{->}(30,0)(42,0)
}

\rput(220,0){
\psframe(0,-5)(18,55)\rput(9,30){$\bA^{-1}$}}

\rput(238,0){
\psline{->}(0,50)(10,50)\rput(15,50){$\hat{\bx}_1$}
\rput(0,30){\rput(13,2){$\vdots$}}
\psline{->}(0,0)(10,0)\rput(15,0){$\hat{\bx}_K$}
}

\rput(169,72){${\mathcal{D}}$}
\psframe[linestyle=dashed](101,-10)(242,68)

\end{pspicture}
\end{center}
\caption{A schematic overview of the integer-forcing source coding framework with the nested lattice pair $\Lambda\subset\Lambda_f$. Each encoder adds a dither $\bd_k$ uniformly distributed over the Voronoi region of the fine lattice $\Lambda_f$ and statistically independent of all other quantities, quantizes the dithered signal onto $\Lambda_f$ and reduces the result modulo the coarse lattice $\Lambda$. The encoding rate is $\frac{1}{n}\log(\Vol(\Lambda)/\Vol(\Lambda_f))$. The decoder subtracts back the dithers and reduces the results modulo $\Lambda$ (this $\Mod$ reduction is actually not necessary and is only illustrated for didactic purposes). Then, the decoder multiplies the signals by a full-rank integer matrix $\bA\in\ZZ^{K\times K}$, reduces the results $\Mod$ and multiplies by $\bA^{-1}$ to form the estimates $\hat{\bx}_1,\ldots,\hat{\bx}_K$.}
\label{fig:IFSCarchitecture}
\end{figure*} 

An important feature of the proposed approach is that it allows the system designer to trade-off performance and complexity. At one extreme, integer-forcing (IF) source coding can be implemented using high-dimensional nested lattices that have near-optimum quantization and channel coding performance. At the other extreme, IF source coding can be implemented with the low-complexity one-dimensional scaled integer lattice $\ZZ$, used as a quantizer as well as a channel code. Surprisingly, the rate loss from using the $1D$ lattice rather than ``good'' high-dimensional nested lattices, amounts to about $2$ bits per sample per encoder, at any distortion level. At high resolution, where the compression rate is high, this loss of $2$ bits is insignificant.

Implementing the $1D$ version of IF source coding only requires each encoder to reduce its observation modulo the lattice $2^R\Delta\ZZ$ and then quantize the obtained signal onto $\Delta\ZZ$, for some $\Delta>0$ which depends on the required distortion. This simple operation can actually be implemented using an analog-to-digital converter (ADC).\footnote{The analog modulo operation is actually already implemented, to some extent, in a class of ADCs called \emph{folding ADCs}~\cite{vp92}.}
The observation that at high resolution $1D$ IF source coding does not lose much w.r.t. the asymptotic performance achieved by Berger-Tung's compression may challenge the current paradigm of ADC design - rather than sample each source at a high rate and then compress it, why not sample at the compression rate to begin with?
An idea in a similar spirit lies at the heart of compressed sensing~\cite{Donoho06}, where the \emph{number of samples} required to reconstruct a sparse signal is reduced according to its sparseness level. Here, the \emph{number of sampled bits} required for reconstructing a source is reduced towards the source's rate-distortion function. The power consumption of an ADC depends on the number of bits it produces per second~\cite{Walden99}.
If the front end of the ADC includes an analog modulo operation, the ADC will need less quantization levels, i.e., less bits. Thus, if analog modulo reduction can be implemented efficiently, the IF approach may potentially lead to a more efficient ADC architectures.

IF source coding can be seen as the source coding dual of IF equalization~\cite{zneg12IT}. IF equalization is a low complexity receiver architecture for the Gaussian MIMO channel. The IF receiver first decodes integer linear combinations of the transmitted codewords, which is possible if all transmitted codewords were taken from the same linear code~\cite{ng11IT}, and then solves these linear combinations for the transmitted codewords. In IF source coding, all encoders first quantize their observations to the desired distortion level, and then reduce them modulo the same lattice $\Lambda$.\footnote{If the quantization is performed by the $1D$ lattice $\Lambda_f=\Delta\ZZ$ and the coarse lattice used for binning is $\Lambda=2^R\Delta\ZZ$, where $2^R$ is a positive integer, the order of the modulo and quantization operations can be switched.} The decoder receives the quantized modulo reduced signals. In order to form estimates of the original signals with the desired distortion level, it has to figure out what was the effect of the modulo reduction on each observation. Rather than doing this directly, it first tries to figure out what is the effect of reducing $K$ linear combinations with integer-valued coefficients of the original signals modulo $\Lambda$, and only then extract the desired effects. See Figure~\ref{fig:IFSCarchitecture}.

The rest of the paper is organized as follows. In Section~\ref{sec:problemstatement} we formally define the distributed lossy compression problem at hand, and introduce the performance benchmark we use throughout the paper which is based on the Berger-Tung inner bound. Basic lattice definitions and figures of merit are recalled in Section~\ref{sec:LQ}, where standard results on lattice quantization are also reviewed. The IF source coding scheme is presented in Section~\ref{sec:IFSC}, and the performance limits of the scheme are derived for the asymptotic case of high-dimensional ``good'' nested lattice codebooks. In Section~\ref{sec:exandapps}, a comparison between the performance of IF source coding and other known coding schemes is given for several scenarios. Applications of IF source coding to several communication problems that are not restricted to pure lossy compression are also given. In particular, we study the performance of a compress-and-forward scheme for relay networks where the compression is performed via IF source coding. We also study the problem of distributively  transmitting $K$ correlated Gaussian random variables over $K$ parallel AWGN channels, and show that IF source coding can improve over standard approaches. In Section~\ref{sec:oneshot} we describe and analyze the one-shot version of IF source coding, where the scaled $1D$ integer lattice is used for quantization and channel coding.

\textit{Notation.} We denote scalars by lowercase letters, vectors by boldface lowercase letters and matrices by boldface uppercase letters, e.g., $x$, $\bx$ and $\bX$. Column vectors usually represent the spatial dimension whereas row vectors represent the time dimension. For example $\bx=[x_1 \ \cdots \ x_K]^T\in\RR^{K\times 1}$ may represent a Gaussian vector of correlated random variables, whereas $\bx_k\in\RR^{1\times n}$ may represent $n$ i.i.d. realizations of the random variable $x_k$. We denote the Euclidean norm of a vector by $\|\cdot \|$ and the absolute value of the determinant of a square matrix by $|\cdot|$.
All variables in the paper are real-valued and all logarithms are to the base $2$.

\section{Problem statement}
\label{sec:problemstatement}
We consider a distributed source coding setting with $K$ encoding terminals and one decoder. Each of the $K$ encoders has access to a vector $\bx_k\in\RR^n$ of $n$ i.i.d. realizations of the random variable $x_k$, $k=1,\ldots,K$. The random vector $\bx=[x_1 \ \cdots \ x_K]^T$ is assumed Gaussian with zero mean and covariance matrix
\begin{align}
\bK_{\bx\bx}\triangleq\mathbb{E}(\bx\bx^T).\nonumber
\end{align}
Each encoder maps its observation $\bx_k$ to an index using the encoding function
\begin{align}
\mathcal{E}_k \ : \ \RR^n\rightarrow \ \{1,\ldots,2^{nR_k}\},\nonumber
\end{align}
and sends the index to the decoder.

The decoder is equipped with $K$ decoding functions
\begin{align}
\mathcal{D}_k \ : \ \{1,\ldots,2^{nR_1}\}\times\cdots\times\{1,\ldots,2^{nR_K}\} \ \rightarrow \ \RR^n,\nonumber
\end{align}
for $k=1,\ldots,K$.
Upon receiving $K$ indices, one from each terminal, the decoder generates estimates
\begin{align}
\hat{\bx}_k=\mathcal{D}_k\left(\mathcal{E}_1(\bx_1),\ldots,\mathcal{E}_K(\bx_K) \right), \ \ k=1,\ldots,K\nonumber.
\end{align}
A rate-distortion vector $(R_1,\ldots,R_K,d_1,\ldots,d_K)$ is achievable if there exist encoding functions $\mathcal{E}_1,\ldots,\mathcal{E}_K$ and decoding functions $\mathcal{D}_1,\ldots,\mathcal{D}_K$ such that
\begin{align}
\frac{1}{n}\mathbb{E}\left(\|\bx_k-\hat{\bx}_k\|^2\right)\leq d_k,\label{distcond}
\end{align}
for all $k=1,\ldots,K$. Let $\bX\triangleq[\bx_1^T \ \cdots \ \bx_K^T]^T$. A conditionally \emph{unbiased} rate-distortion vector $(R_1,\ldots,R_K,d_1,\ldots,d_K)$ is achievable if in addition to~\eqref{distcond}, the condition
\begin{align}
\mathbb{E}(\hat{\bx}_k|\bX)=\bx_k, \ \ k=1,\ldots,K\label{biascond}
\end{align}
is satisfied for any realization of $\bX$. Note that this condition is equivalent to
\begin{align}
\mathbb{E}(\bx_k-\hat{\bx}_k|\bX)=0 , \ \ k=1,\ldots,K.\nonumber
\end{align}
Although condition~\eqref{biascond} is not as common in the literature as condition~\eqref{distcond}, in this paper we restrict attention to the conditionally unbiased case, i.e., we impose condition~\eqref{biascond}. Several applications of interest require the estimates formed by the decoder to be conditionally unbiased. For instance, consider a communication scenario where distributed antenna terminals observe noisy linear combinations of the signals transmitted by several encoders and want to forward a compressed version of these signals to a central processor that needs to decode the transmitted messages. In such a scenario it is most convenient to treat the quantization noise as an additive one, meaning that it is statistically independent of the signals that are being quantized. This amounts to requiring condition~\eqref{biascond}. Moreover, when the conditionally unbiased requirement~\eqref{biascond} is not essential to the application at hand, one can always perform minimum mean-squared estimation of $\bX$ from $\hat{\bX}$ and further reduce the MSE distortion.

We further focus on the symmetric case where $R_1=\cdots=R_K=R$ and $d_1=\cdots=d_k=d$. We do this for three reasons.
First, such a symmetry constraint naturally arises in many applications, where the coding burden has to be equally split between the distributed encoders. Second, this allows for a simpler description of the proposed coding scheme and the rate-distortion region it achieves. Finally, in an asymmetric setting there exist several examples where structured binning outperforms the standard approach of Berger-Tung compression~\cite{ng07IT,kp09,wagner11}. 
Focusing on the symmetric case eliminates the possibility of such examples that are, to some extent, skewed towards using structured binnining. Nevertheless, we stress that the scheme proposed in this paper is not restricted to the symmetric case, and can be easily extended to achieve asymmetric rate-distortion vectors by using a more complicated \emph{chain} of nested lattices, rather than the nested lattice \emph{pair} we use in the sequel.

Finding the full rate-distortion region, i.e., the set of all achievable rate-distortion vectors, for the described setup is an open problem for $K>2$. For $K=2$, Wagner \textit{et al}.~\cite{wtv08} showed that the Berger-Tung approach is optimal. This approach consists of quantizing each source using standard single-source rate-distortion theory with a Gaussian test channel, and then using Slepian-Wolf encoding for compressing the quantization indices. For $K>2$ it is now known that the Berger-Tung approach does not attain the full rate-distortion region (see e.g.~\cite{kp09}). However, to the best of our knowledge, it is not known whether the Berger-Tung inner bound is loose for the symmetric case. In the absence of a better known coding scheme, we take the symmetric rate from Berger-Tung's inner bound as our benchmark. More specifically, the sum-rate in Berger-Tung's inner bound is given by
\begin{align}
\sum_{k=1}^K R_k\geq I(\bx;\bu),\label{BTsum}
\end{align}
where $\bu=[u_1 \ \cdots u_K]^T$ is a vector of auxiliary random variables that satisfy the set of Markov chains
\begin{align}
u_k-x_k-\left(\{x_j,u_j\}_{j\neq k}\right)\nonumber
\end{align}
and such that there exist functions $\hat{x}_k(u_1,\ldots,u_K)$ satisfying $\mathbb{E}(x_k-\hat{x}_k)^2<d_k$ for all $k=1,\ldots,K$.
Optimizing over $\bu$ is a difficult task. A common and natural choice in the quadratic-Gaussian case is taking
\begin{align}
u_k=\alpha_k x_k+w_k, \ \ k=1,\ldots,K,\label{testchannel}
\end{align}
where $w_1,\ldots,w_K$ are statistically independent zero mean Gaussian random variables that are also independent of $\bx$, and $\alpha_1,\ldots,\alpha_K$ are some constants~\cite{tvw10}. Such a choice was shown to be optimal for $K=2$~\cite{wtv08}, but may be suboptimal for larger dimensions. Since we are after conditionally unbiased estimates for the $K$ components of $\bx$, we set $\alpha_k=1$, $w_k\sim\mathcal{N}(0,d)$ and $\hat{x}_k(u_1,\ldots,u_K)=u_k$ for all $k=1,\ldots,K$. Substituting this choice in~\eqref{BTsum} gives
\begin{align}
\sum_{k=1}^K R_k&\geq \frac{1}{2}\log\frac{|\bK_{\bx\bx}+d\bI|}{|d\bI|}\nonumber\\
&=\frac{1}{2}\log \left|\bI+\frac{1}{d}\bK_{\bx\bx} \right|.\label{benchmarktmp}
\end{align}
This sum-rate is achievable using Berger-Tung compression. In this paper we are interested in the symmetric rate-distortion region. To this end, we take~\eqref{benchmarktmp} normalized by $K$ as our benchmark
\begin{align}
R^{\text{BT}}_{\text{bench}}(d)\triangleq\frac{1}{2K}\log\left|\bI+\frac{1}{d}\bK_{\bx\bx} \right|.\label{benchmark}
\end{align}
Note that $R^{\text{BT}}_{\text{bench}}(d)$ is not a lower bound on the minimal symmetric rate-distortion function achieved by Berger-Tung compression, as our choice of $\bu$ is not necessarily the best one. It is also not an upper bound on the minimal symmetric rate-distortion function achieved by Berger-Tung compression, as the symmetric rate with our choice of $\bu$ may not be dominated by the sum-rate constraint. 

\section{Preliminaries}
\label{sec:LQ}
In this section we recall several lattice properties that will be useful in the sequel and review the concept of dithered lattice quantization.

A lattice $\Lambda$ is a discrete subgroup of $\RR^n$ which is closed under reflection and real addition. We denote the nearest neighbor quantizer associated with the lattice $\Lambda$ by
\begin{align}
\Ql(\by)=\argmin_{\mathbf{t}\in\Lambda}\|\by-\mathbf{t}\|.
\label{NNquantizer}
\end{align}
The basic Voronoi region of $\Lambda$, denoted by $\CV$, is the set of all points in $\RR^n$ which are quantized to the zero vector, where ties in~\eqref{NNquantizer} are broken in a systematic manner. The modulo operation returns the quantization error w.r.t. the lattice,
\begin{align}
\left[\by\right]\Mod=\by-\Ql(\by)\nonumber
\end{align}
and satisfies the property
\begin{align}
[a[\by]\Mod]\Mod=[a\by]\Mod\label{modprop}
\end{align}
for any $a\in\ZZ$ and $\by\in\RR^n$. This property will be used extensively in the sequel.
The second moment of $\Lambda$ is defined as
\begin{align}
\sigma^2(\Lambda)=\frac{1}{n}\frac{1}{\Vol(\CV)}\int_{\bu\in\CV}\|\bu\|^2d\bu,\nonumber
\end{align}
where $\Vol(\CV)$ is the volume of $\CV$. The effective radius of a lattice $r_{\text{eff}}(\Lambda)$ is defined as the radius of an $n$-dimensional ball whose volume equals $\Vol(\CV)$.

The lattice $\Lambda$ can be used for quantizing continuous sources. In particular, an encoder which is interested in conveying a source $\by\in\RR^n$ to a decoder can compute $\Ql(\by)$, which is a lattice point in $\Lambda$, and send a description of this point to the decoder. The quantization error of such a scheme is $\be=\by-\Ql(\by)$, which is a deterministic function of $\by$. Recall that in this paper we are interested in encoder/decoder pairs that produce conditionally unbiased estimates of the source, which is clearly not the case for a standard lattice quantizer. This may be overcome by allowing the encoder and decoder to use common randomness. Let $\bd\sim\Unif(\CV)$ be a random dither vector uniformly distributed over $\CV$ and statistically independent of $\by$, known to both the encoder and the decoder. The \emph{dithered lattice qunatizer} associated with the lattice $\Lambda$ computes $\Ql(\by+\bd)$ and sends a description of the obtained lattice point to the decoder. The decoder produces the estimate
\begin{align}
\hat{\by}&=\Ql(\by+\bd)-\bd\nonumber\\
&=\by+\Ql(\by+\bd)-(\by+\bd)\nonumber\\
&=\by-[\by+\bd]\Mod.\nonumber
\end{align}
The Crypto Lemma~\cite[Lemma 1]{ez04} ensures that the estimation error $-[\by+\bd]\Mod$ is statistically independent of $\by$ and is uniformly distributed over $\CV$. The symmetry of the Voronoi region $\CV$ guarantees that the estimation error has the same distribution as $\bd$ and has zero mean. Thus, $\hat{\by}=\by+\bd$ in distribution, and is a conditionally unbiased estimate of $\by$. Clearly, the average MSE distortion attained by dithered lattice quantization is given by
\begin{align}
\frac{1}{n}\mathbb{E}\left(\|\by-\hat{\by}\|^2\right)=\frac{1}{n}\mathbb{E}(\|\bd\|^2)=\sigma^2(\Lambda).\nonumber
\end{align}
Of course, dithered lattice quantization, as described above, requires an infinite rate as there is an infinite number of points in $\Lambda$. This can be handled using an entropy coded dithered quantizer (ECDQ)~\cite{ziv85,zf92,ramibook}, or a nested lattice codebook~\cite{zse02}. In this work we take the latter approach.

The following definitions characterize the lattice ``goodness'' properties needed in this paper.

\begin{definition}[Goodness for MSE quantization]
\label{def:MSEgoodness}
A lattice $\Lambda$, or more precisely, a sequence of lattices with growing dimension $n$, is said to be good for MSE quantization if\footnote{Note that our condition for MSE goodness is equivalent to the more commonly used condition $\sigma^2(\Lambda)/\Vol(\CV)^{2/n}\rightarrow 1/(2\pi e)$ since the volume of a unit $n$-dimensional ball grows like $(2 \pi e/n)^{n/2}$.}
\begin{align}
\lim_{n\rightarrow\infty} \sigma^2(\Lambda)=\lim_{n\rightarrow\infty}\frac{r^2_{\text{eff}}(\Lambda)}{n}.\nonumber
\end{align}
\end{definition}

\vspace{1mm}

\begin{definition}[Semi-norm ergodic noise]
\label{def:seminormergodic}
We say that a random noise vector $\bz$, or more precisely, a sequence of random noise vectors with growing dimension $n$, with (finite) effective variance \mbox{$\sigma^2_{\bZ}\triangleq\mathbb{E}\|\bz\|^2/n$}, is \emph{semi norm-ergodic} if for any $\epsilon>0$, $\delta>0$ and $n$ large enough
\begin{align}
\Pr\left(\|\bz\|>\sqrt{(1+\delta)n\sigma^2_{\bz}} \right)\leq\epsilon.\label{normergodicDef}
\end{align}
Note that by the law of large numbers, any i.i.d. noise is semi norm-ergodic.
\end{definition}

\vspace{1mm}

The next Lemma restates Corollary 2 from~\cite{oe12} to fit our purposes.

\vspace{1mm}

\begin{lemma}
\label{lem:mixturenoise}
Let \mbox{$\bd_1,\cdots,\bd_K$} be statistically independent random dither vectors, each uniformly distributed over the Voronoi region $\CV$ of a lattice $\Lambda$ that is good for MSE quantization. Let $\bz$ be an i.i.d. random vector statistically independent of \mbox{$\left\{\bd_1,\cdots,\bd_K\right\}$}. Any deterministic linear combination of $\bd_1,\cdots,\bd_K,\bz$ is semi norm-ergodic.
\end{lemma}

\vspace{1mm}

\begin{definition}[Goodness for channel coding]
\label{def:codinggoodness}
A lattice $\Lambda$, or more precisely, a sequence of lattices with growing dimension $n$, is said to be good for channel coding if for any \mbox{$0<\delta<1$} and any $n$-dimensional semi norm-ergodic vector $\bz$ with zero mean and  effective variance \mbox{$\mathbb{E}\|\bz\|^2/n<(1-\delta)r^2_{\text{eff}}(\Lambda)/n$}
\begin{align}
\lim_{n\rightarrow\infty} \Pr\left(\bz\notin\CV \right)=0.\nonumber
\end{align}
\end{definition}

\vspace{2mm}

A lattice $\Lambda$ is said to be nested in $\Lambda_f$ if $\Lambda\subseteq\Lambda_f$. The coding scheme presented in this paper utilizes a pair of nested lattices such that the fine lattice $\Lambda_f$ is good for MSE quantization and the coarse lattice $\Lambda$ is good for channel coding. An ensemble for drawing pairs of nested lattices that satisfy these goodness properties is described in~\cite{oe12},\footnote{In~\cite{oe12} the definition of goodness for channel coding was weaker than that needed here. In particular, only the existence of lattices that achieve a vanishing error probability under \emph{coset} nearest neighbor decoding in the present of semi-norm ergodic noise was proved. However, a more careful inspection of the derivation in~\cite{oe12} reveals that the probability of decoding an erroneous point in the correct coset also vanishes with the dimension $n$ for the choice of lattice parameters made in~\cite{oe12}. Thus, the existence of pairs of nested lattices such that both fine and coarse lattices are good for MSE quantization and channel coding follows.} and the existence of lattice pairs with slightly more demanding ``goodness'' requirements was shown in~\cite{elz05,kp09}. A nested lattice code $\mathcal{C}=\Lambda_f\cap\CV$ with rate
\begin{align}
R=\frac{1}{n}\log\left(\frac{\Vol(\Lambda)}{\Vol(\Lambda_f)}\right)=\frac{1}{2}\log\left(\frac{r^2_{\text{eff}}(\Lambda)}{r^2_{\text{eff}}(\Lambda_f)}\right)
\end{align}
is associated with the nested lattice pair.

Before describing the integer-forcing source coding scheme, let us illustrate how the codebook $\mathcal{C}$ described above can be used for compressing $n$ samples of a single memoryless Gaussian source $Y\sim\mathcal{N}(0,P)$ with distortion $d$. Assume that the fine lattice $\Lambda_f$, which is good for MSE quantization, has second moment $\sigma^2(\Lambda_f)=d$. This implies that $r^2_{\text{eff}}(\Lambda_f)/n\rightarrow d$. The coarse lattice $\Lambda$, which is good for AWGN channel coding, has effective radius $r^2_{\text{eff}}(\Lambda)=n(P+d+\epsilon)$, for some arbitrarily small $\epsilon>0$. A dither $\bd$ uniformly distributed over $\CV_f$ is known to both the encoder and the decoder. The encoder computes
\begin{align}
[\Qlf(\by+\bd)]\Mod\in\mathcal{C},\nonumber
\end{align}
and sends its index to the decoder. The decoder computes
\begin{align}
\hat{\by}&=\left[[\Qlf(\by+\bd)]\Mod-\bd\right]\Mod\nonumber\\
&\stackrel{(i.d.)}{=}\left[\by+\bd\right]\Mod\nonumber\\
&\stackrel{(w.h.p.)}{=}\by+\bd\label{whpeq}
\end{align}
where $\stackrel{(i.d.)}{=}$ stands for  equality in distribution and $\stackrel{(w.h.p.)}{=}$ for equality with high probability. The equality~\eqref{whpeq} follows from the fact that the random vector $\by+\bd$ is semi-norm ergodic due to Lemma~\ref{lem:mixturenoise} and has effective variance $\mathbb{E}(\|\by+\bd\|^2)/n=P+d$. Since $\Lambda$ is good for channel coding and $\mathbb{E}\|\by+\bd\|^2/n<r^2_{\text{eff}}(\Lambda)/n$, the probability that $\Ql(\by+\bd)\neq\mathbf{0}$ vanishes, and hence, $\left[\by+\bd\right]\Mod\stackrel{(w.h.p.)}{=}\by+\bd$. Thus, with high probability
\begin{align}
\frac{1}{n}\mathbb{E}(\|\by-\hat{\by}\|^2)=\frac{1}{n}\mathbb{E}(\|\bd\|^2)=d,\nonumber
\end{align}
as desired. The required rate for achieving this distortion is
\begin{align}
R(d)&=\frac{1}{2}\log\left(\frac{r^2_{\text{eff}}(\Lambda)}{r^2_{\text{eff}}(\Lambda_f)}\right)\nonumber\\
&=\frac{1}{2}\log\left(\frac{n(P+d+\epsilon)}{nd}\right)\nonumber\\
&=\frac{1}{2}\log\left(1+\frac{P+\epsilon}{d}\right)
\end{align}
where the additional $+1$ inside the logarithm, w.r.t. the standard Gaussian rate-distortion function, is a consequence of our requirement that the reconstruction $\hat{\by}$ forms a conditionally unbiased estimate of $\by$. In fact, we can eliminate this term by performing an additional Wiener estimation step on $\hat{\by}$, at the expense of introducing bias~\cite{ramibook}.

\section{Integer-Forcing Source Coding}
\label{sec:IFSC}
In the IF distributed source coding scheme all encoders use the same nested lattice codebook $\mathcal{C}=\Lambda_f\cap\CV$, constructed from the nested lattice pair $\Lambda\subset\Lambda_f$, with rate
\begin{align}
R=\frac{1}{2}\log\left(\frac{r^2_{\text{eff}}(\Lambda)}{r^2_{\text{eff}}(\Lambda_f)}\right).\nonumber
\end{align}
As in the previous section, the fine lattice $\Lambda_f$ is good for MSE quantization with $\sigma^2(\Lambda_f)=d$ whereas the coarse lattice $\Lambda$ is good for channel coding. All encoders employ a similar encoding operation. The $k$th encoder uses a dither $\bd_k$, statistically independent of everything else and uniformly distributed over $\CV_f$, and employs dithered quantization of $\bx_k$ onto $\Lambda_f$. Then, it reduces the obtained lattice point modulo the coarse lattice $\Lambda$ and sends $nR$ bits describing the index of the resulting point to the decoder. Specifically, the $k$th encoder conveys the index corresponding to the point
\begin{align}
\left[\Qlf(\bx_k+\bd_k) \right]\Mod\nonumber
\end{align}
to the decoder.

The decoder first subtracts back the dithers from each of the reconstructed signals and reduces the results modulo $\Lambda$, giving rise to
\begin{align}
\tilde{\bx}_k&=\left[\left[\Qlf(\bx_k+\bd_k) \right]\Mod -\bd_k \right]\Mod\nonumber\\
&=\left[\bx_k+\left[\Qlf(\bx_k+\bd_k) \right]\Mod -(\bx_k+\bd_k) \right]\Mod\nonumber\\
&\stackrel{(i.d.)}{=}\left[\bx_k+\bd_k \right]\Mod\label{encmodests}
\end{align}
If the coarse lattice $\Lambda$ is chosen such that its effective radius is large enough, the modulo operation in~\eqref{encmodests} would have no effect on $\bx_k+\bd_k$, and the decoder would have estimates of each $\bx_k$ with average MSE of $d$, as desired. However, the encoding rate grows with $r_{\text{eff}}^2(\Lambda)$, and we would therefore prefer to choose it as small as possible.

The key idea behind IF source coding is that if the elements of $\bx$ are correlated, then linear combinations of $\{\bx_k+\bd_k\}_{k=1}^K$ with integer-valued coefficients may have smaller effective variances than the original signals. The IF decoder therefore first estimates $K$ integer linear combinations of $\{\bx_k+\bd_k\}_{k=1}^K$, and then uses these estimates for estimating the desired signals. Using this approach, $r_{\text{eff}}^2(\Lambda)$ should only be greater than the largest effective variance among the $K$ linear combinations. When the entries of $\bx$ are sufficiently correlated, and the integer-valued coefficients are chosen appropriately, this may significantly reduce the required encoding rate.

\vspace{1mm}

Let $\bX=[\bx_1^T \ \cdots \ \bx_K^T]^T$, $\bD=[\bd_1^T \ \cdots \ \bd_K^T]^T$ and $\tilde{\bX}=[\tilde{\bx}_1^T \ \cdots \ \tilde{\bx}_K^T]^T$. Using this notation, the decoder has access to
\begin{align}
\tilde{\bX}=\left[\bX+\bD\right]\Mod,\nonumber
\end{align}
where the notation $\hspace{-2.5mm}\mod\hspace{-0.2mm}\Lambda$ is to be understood as reducing \emph{each row} of the obtained matrix modulo the coarse lattice. The decoder chooses a full-rank integer-valued matrix $\bA\in\ZZ^{K\times K}$ and computes
\begin{align}
\widehat{\bA\bX}&\triangleq\left[\bA\tilde{\bX}\right]\Mod\nonumber\\
&=\left[\bA\left[\bX+\bD\right]\Mod\right]\Mod\nonumber\\
&=\left[\bA(\bX+\bD)\right]\Mod\label{modeqs}
\end{align}
where~\eqref{modeqs} follows from the modulo property~\eqref{modprop}.

Let $\ba_k^T$ be the $k$th row of the matrix $\bA$. The random vector $\ba_k^T(\bX+\bD)$ satisfies the conditions of Lemma~\ref{lem:mixturenoise} as $\ba_k^T\bX$ is an i.i.d. Gaussian vector and each of the statistically independent dithers $\bd_1,\ldots,\bd_K$ is uniformly distributed over the Voronoi region of a lattice that is good for MSE quantization. Therefore, $\ba_k^T(\bX+\bD)$ is semi-norm ergodic. It follows from the goodness of $\Lambda$ for channel coding that if
\begin{align}
\frac{\mathbb{E}\left(\|\ba_k^T(\bX+\bD)\|^2\right)}{n}<\frac{r_{\text{eff}}^2(\Lambda)}{n}\nonumber
\end{align}
then for $n$ large enough
\begin{align}
\left[\ba_k^T(\bX+\bD)\right]\Mod\stackrel{(w.h.p.)}{=}\ba_k^T(\bX+\bD).\nonumber
\end{align}
Moreover, if this holds for all $k=1,\ldots,K$, i.e., if
\begin{align}
\max_{k=1,\ldots,K}\frac{\mathbb{E}\left(\|\ba_k^T(\bX+\bD)\|^2\right)}{n}<\frac{r_{\text{eff}}^2(\Lambda)}{n}\nonumber
\end{align}
then for $n$ large enough
\begin{align}
\widehat{\bA\bX}\stackrel{(w.h.p.)}{=}\bA(\bX+\bD).\label{inteqwhp}
\end{align}
Noting that
\begin{align}
\frac{\mathbb{E}\left(\|\ba_k^T(\bX+\bD)\|^2\right)}{n}=\ba_k^T(\bK_{\bx\bx}+d\bI)\ba_k,\nonumber
\end{align}
this implies that for~\eqref{inteqwhp} to hold, it suffices to set
\begin{align}
\frac{r_{\text{eff}}^2(\Lambda)}{n}=\max_{k=1,\ldots,K}\ba_k^T(\bK_{\bx\bx}+d\bI)\ba_k +\epsilon\nonumber
\end{align}
for some arbitrarily small $\epsilon>0$, which corresponds to a rate of
\begin{align}
R=\frac{1}{2}\log\left(\frac{\max_{k=1,\ldots,K}\ba_k^T(\bK_{\bx\bx}+d\bI)\ba_k +\epsilon}{d}\right).\nonumber
\end{align}
The decoder proceeds by computing
\begin{align}
\hat{\bX}=\bA^{-1}\widehat{\bA\bX}\stackrel{(w.h.p.)}{=}\bX+\bD,\nonumber
\end{align}
which is (w.h.p.) a conditionally unbiased estimate of $\bX$ with average MSE distortion $d$ per component. The next theorem summarizes the performance of IF source coding.

\vspace{2mm}

\begin{theorem}[Performance of IF source coding]
\label{thm:IFrd}
For any distortion $d>0$ and any choice of $\bA=[\ba_1 \ \cdots \ \ba_K]^T\in\ZZ^{K\times K}$, there exists a (sequence of) nested lattice pair(s) $\Lambda\subset\Lambda_f$ such that IF source coding can achieve any rate satisfying
\begin{align}
R>R_{\text{IF}}(\bA,d)\triangleq\frac{1}{2}\log\left(\max_{k=1,\ldots,K}\ba_k^T\left(\bI+\frac{1}{d}\bK_{\bx\bx}\right)\ba_k \right).\nonumber
\end{align}
For the optimal choice of $\bA$, IF source coding can achieve any rate satisfying
\begin{align}
R>R_{\text{IF}}(d)\triangleq\frac{1}{2}\log\left(\min_{\substack{{\bA\in\ZZ^{K\times K}}\\ {\det(\bA)\neq 0}}}\max_{k=1,\ldots,K}\ba_k^T\left(\bI+\frac{1}{d}\bK_{\bx\bx}\right)\ba_k \right).\nonumber
\end{align}
\end{theorem}

\vspace{5mm}

The matrix $\bI+\frac{1}{d}\bK_{\bx\bx}$ is symmetric and positive definite, and therefore it admits a Cholesky decomposition
\begin{align}
\bI+\frac{1}{d}\bK_{\bx\bx}=\bF\bF^T,\label{chol}
\end{align}
where $\bF$ is a lower triangular matrix with strictly positive entries. With this notation,
\begin{align}
R_{\text{IF}}(d)=\frac{1}{2}\log\left(\min_{\substack{{\bA\in\ZZ^{K\times K}}\\ {\det(\bA)\neq 0}}}\max_{k=1,\ldots,K}\|\bF \ \ba_k\|^2 \right).\label{RIFdtmp}
\end{align}
Denote by $\Lambda(\bF^T)$ the $K$ dimensional lattice spanned by the matrix $\bF^T$, i.e.,
\begin{align}
\Lambda(\bF^T)\triangleq\left\{\bF^T\ba \ : \ \ba\in\ZZ^{K} \right\}.\nonumber
\end{align}
It follows that the problem of finding the optimal matrix $\bA$ is equivalent to finding the $K$ shortest linearly independent vectors of $\Lambda(\bF^T)$. Although this problem is NP-hard in general, its solution can be efficiently approximated using the LLL algorithm~\cite{lll82}, whose running time is polynomial.

Moreover, we can express the rate-distortion function achieved by IF source coding using the \emph{successive minima} of the lattice $\Lambda(\bF^T)$.

\vspace{1mm}

\begin{definition}[Successive minima]
\label{def:sucmin}
Let $\Lambda(\mathbf{G})$ be the lattice spanned by the full-rank matrix $\mathbf{G} \in \mathbb{R}^{K \times K}$. For \mbox{$k=1,\ldots,K$}, we define the $k$th successive minimum as
\begin{align}
\lambda_k(\mathbf{G}) \triangleq \inf\left\{r \ : \ \dim\left(\Span\left(\Lambda(\mathbf{G})\bigcap \mathcal{B}(\mathbf{0},r)\right)\right)\geq k\right\}\nonumber
\end{align}
where $\mathcal{B}(\mathbf{0},r)=\left\{\bx\in\RR^{K} \ : \ \|\bx\|\leq r\right\}$ is the closed ball of radius $r$ around $\mathbf{0}$. In words, the $k$th successive minimum of a lattice is the minimal radius of a ball centered around $\mathbf{0}$ that contains $k$ linearly independent lattice points.
\end{definition}

\vspace{1mm}

Using Definition~\ref{def:sucmin} and~\eqref{RIFdtmp}, the IF rate-distortion function is given by
\begin{align}
R_{\text{IF}}(d)=\frac{1}{2}\log\left(\lambda^2_K(\mathbf{\bF^T})\right),\label{RIFd}
\end{align}
where the dependence of the r.h.s. on $d$ is through the matrix $\bF$ defined in~\eqref{chol}.

Next, we show in Lemma~\ref{lem:IFvsBT} that the performance of IF source coding, in the symmetric setting considered, is inferior to the Berger-Tung benchmark, i.e.,
$R_{\text{IF}}(d)\geq R^{\text{BT}}_{\text{bench}}(d)$. We will need the simple following proposition.

\vspace{1mm}

\begin{proposition}
\label{prop:prodineq}
For a lattice spanned by some full rank matrix $\bG\in\RR^{K\times K}$
\begin{align}
|\bG|\leq \prod_{k=1}^K \lambda_k(\mathbf{G})\nonumber
\end{align}
\end{proposition}

\vspace{1mm}

\begin{proof}
Let $\ba_1,\ldots,\ba_K\in\ZZ^K$ be $K$ linearly independent vectors such that $\lambda_k(\bG)=\|\bG\ba_k\|$ for all $k=1,\ldots,K$, and let $\bA=[\ba_1 \ \cdots \ \ba_K]$. Since all entries of $\bA$ are integer-valued we must have $|\bA|\geq 1$, and therefore
\begin{align}
|\bG|&\leq|\bG| \ |\bA|=|\bG\bA|\nonumber\\
&=\left|[\bG\ba_1 \ \cdots \ \bG\ba_K]\right|\leq \prod_{k=1}^K \|\bG\ba_k\|\nonumber\\
&=\prod_{k=1}^K \lambda_k(\bG).\nonumber
\end{align}
\end{proof}

\vspace{1mm}

\begin{lemma}
\label{lem:IFvsBT}
For any $d>0$ and for any choice of full-rank $\bA\in\ZZ^{K\times K}$ we have
\begin{align}
\frac{1}{2}\log\left(\max_{k=1,\ldots,K}\ba_k^T(\bI+\frac{1}{d}\bK_{\bx\bx})\ba_k \right)\geq \frac{1}{2K}\log \left|\bI+\frac{1}{d}\bK_{\bx\bx} \right|,\nonumber
\end{align}
and therefore, in the considered symmetric setting, the rate-distortion function $R_{\text{IF}}(d)$ of IF source coding is never smaller than the benchmark $R^{\text{BT}}_{\text{bench}}(d)$.
\end{lemma}

\vspace{1mm}

\begin{proof}
Let $\bF$ be as defined in~\eqref{chol}.
For the optimal choice of $\bA$ and for any $d>0$ we have
\begin{align}
\frac{1}{2}\log\bigg(\max_{k=1,\ldots,K}\ba_k^T(\bI+&\frac{1}{d}\bK_{\bx\bx})\ba_k \bigg)= \frac{1}{2}\log\left(\lambda^2_K(\mathbf{\bF^T})\right)\nonumber\\
&\geq\frac{1}{2}\frac{1}{K}\sum_{k=1}^K\log\left(\lambda^2_k(\mathbf{\bF^T})\right)\label{monotonicity}\\
&=\frac{1}{2K}\log\left(\prod_{k=1}^K\lambda^2_k(\mathbf{\bF^T})\right)\nonumber\\
&\geq\frac{1}{2K}\log\left(|\bF|^2\right)\label{prodineq}\\
&=\frac{1}{2K}\log \left|\bI+\frac{1}{d}\bK_{\bx\bx} \right|\label{deteq},
\end{align}
where~\eqref{monotonicity} follows from the monotonicity of $\lambda_k(\mathbf{\bF^T})$ in $k$ along with the monotonicity of the logarithm function,~\eqref{prodineq} follows from Proposition~\ref{prop:prodineq} and~\eqref{deteq} follows from~\eqref{chol}.
\end{proof}

\vspace{1mm}

As discussed in Section~\ref{sec:problemstatement}, in an asymmetric problem setting, structured binning may result in a better rate-distortion region than the one obtained by Berger-Tung compression. Lemma~\ref{lem:IFvsBT} shows that under the symmetric setup, at least with IF source coding, this may not be the case. Nevertheless, the complexity reduction obtained by using IF source coding rather than Berger-Tung compression makes it an attractive candidate for practical implementation. Moreover, as we shall see in Section~\ref{sec:oneshot}, a one-shot version of IF source coding can be easily derived and analyzed. Although one-shot versions of Berger-Tung compression were also considered in~\cite{yag13} and an inner bound was derived, it is unclear how to interpret this inner bound for the problem at hand.

\begin{remark}
The crucial element in the IF source coding scheme is that all encoders reduce their quantized signals modulo the same coarse lattice. The modulo reduction plays the role of binning. Theoretically, each encoder can first reduce its observation $\Mod$ and only then quantize it using a quantizer designed for the modulo reduced source~\cite{ramibook}. No nesting is required between the quantizer and the coarse lattice.
This results in the decoder receiving the signals \mbox{$\tilde{\bx}_k=[\bx_k]\Mod+\bd_k$}, $k=1,\ldots,K$, where $\bd_k$ is quantization noise. The decoder can proceed to compute $\widehat{\bA\bX}$ as described above. The difficulty with such an implementation is that the quantizer needs to be matched to the modulo reduced source, which requires some sort of (high-dimensional) entropy coding. As we shall see in Section~\ref{subsec:modADCs}, in the $1D$ version of IF source coding, where the coarse lattice as well as the quantizer are scaled integer lattices, the modulo reduction can precede quantization without increasing complexity.
\end{remark}

\begin{remark}
Another implementation issue to consider is the goodness requirements on $\Lambda$. When $\Lambda$ is used for modulation over the AWGN channel, it suffices to require that $\Lambda$ is good under \emph{coset} nearest neighbor decoding. This means that $\Lambda$ is split into cosets, usually using a coarse lattice nested inside it, and the decoder only needs to choose the coset the transmitted point belongs to. As a result, when coding for the AWGN channel is considered, a construction A lattice~\cite{cs88,loeliger97} with a linear codebook of small prime cardinality $p$ suffices to achieve a vanishing error probability. In such a construction, the minimum distance is limited by $p$, and the error probability for decoding the actual point transmitted, rather than the coset, cannot vanish with the dimension. However, all pairs of points with nonincreasing (as a function of $n$, the code dimension) Euclidean distance belong to the same coset, and therefore such a lattice is still good for coset nearest neighbor decoding.

In IF source coding, the decoder needs to decode the \emph{actual} lattice point of $\Lambda$ closest to $\ba_k^T(\bX+\bD)$, rather than just its coset. Therefore, construction A lattices obtained from a linear codebook with small $p$ do not suffice in order to achieve a vanishing error probability. However, one can still achieve a very small error probability, though not vanishing with the dimension, using standard Construction A lattices with moderate values of $p$. See Section~\ref{sec:oneshot} for further discussion of implementation issues.
\end{remark}

\section{Examples and applications}
\label{sec:exandapps}

This section provides several examples that demonstrate the performance of IF source coding, along with applications and communication scenarios where IF source coding is advantageous. The section consists of three parts. First we compare the performance of IF source coding to that of a naive distributed compression scheme that ignores the correlation between the sources and to the Berger-Tung benchmark. Then, we use IF source coding as a building block in a Gaussian layered relay network, and demonstrate its advantages compared to other known low complexity schemes. Finally, we show how the idea behind IF source coding can be extended to form a signal-to-noise ratio (SNR) independent joint source channel coding scheme, whose distortion decreases as the SNR improves.

\subsection{Examples}
\label{subsec:examples}

In this subsection we evaluate the minimal symmetric rate needed in order to achieve a conditionally unbiased average MSE of $d$ for two schemes:
\begin{enumerate}
\item IF source coding - this rate is given in Theorem~\ref{thm:IFrd}.
\item Compressing each source using standard rate-distortion theory without exploiting the correlations between the sources - this rate is given by
\begin{align}
R_{\text{naive}}(d)=\max_{k=1,\ldots,K}\frac{1}{2}\log\left(1+\frac{\bK_{\bx\bx}(k,k)}{d}\right),\label{Rdnaive}
\end{align}
and is identical to the rate obtained using IF source coding with the choice $\bA=\bI$.
\end{enumerate}
We also compare these rates to the Beger-Tung benchmark $R^{\text{BT}}_{\text{bench}}(d)$~\eqref{benchmark}.

\begin{example}[Integer decomposable covariance matrix]
As a first example, consider the case where $\bx$ is a Gaussian source with zero mean and covariance matrix $\bK_{\bx\bx}=\bB^{-1}\bB^{-T}$ for some full-rank integer matrix $\bB\in\ZZ^{K\times K}$ with determinant $|\bB|=1$.

The Berger-Tung benchmark symmetric rate-distortion function is given by
\begin{align}
R^{\text{BT}}_{\text{bench}}(d)&=\frac{1}{2K}\log \left|\bI+\frac{1}{d}\bB^{-1}\bB^{-T} \right|\nonumber\\
&=\frac{1}{2K}\left(\log|\bB|^{-2}+\log|\bB\bB^T+\frac{1}{d}\bI| \right)\nonumber\\
&=\frac{1}{2K}\log|\bB\bB^T+\frac{1}{d}\bI|. \nonumber
\end{align}
It can be seen that $R^{\text{BT}}_{\text{bench}}(d)\rightarrow -1/2\log(d)$ as $d\rightarrow 0$.

For IF source coding, one can choose $\bA=\bB$. This choice gives
\begin{align}
R_{\text{IF}}(\bB,d)&=\frac{1}{2}\log\left(\max_{k=1,\ldots,K}\bb_k^T\left(\bI+\frac{1}{d}\bK_{\bx\bx}\right)\bb_k \right)\nonumber\\
&=\frac{1}{2}\log\left(\max_{k=1,\ldots,K}\|\bb_k\|^2+\frac{1}{d} \right)\nonumber
\end{align}
It is easy to see that $R_{\text{IF}}(\bB,d)\rightarrow -1/2\log(d)$ as $d\rightarrow 0$, just as the benchmark rate-distortion function, and therefore, according to Lemma~\ref{lem:IFvsBT}, the choice $\bA=\bB$ is optimal at high resolution.

The naive approach that compresses each source without exploiting the existing correlations fails to achieve the benchmark rate-distortion function. In fact, it can only achieve
\begin{align}
R_{\text{naive}}(d)=\frac{1}{2}\log\left(1+\frac{\max_{k=1,\ldots,K}\|\tilde{\bb}_k\|^2}{d}\right),
\end{align}
where $\tilde{\bb}_k^T$ is the $k$th row of $\bB^{-1}$. All entries of $\tilde{\bb}_k$ are integer-valued since the matrix $\bB$ is integer-valued with determinant $1$. Therefore $\|\tilde{\bb}_k\|^2\geq 1$ for all $k=1,\ldots,K$.
The obtained compression rate approaches $\frac{1}{2}\log(\max\|\tilde{\bb}_k\|^2)-1/2\log(d)$ as $d\rightarrow 0$. Thus, at high resolution, IF source coding requires $\frac{1}{2}\log(\max\|\tilde{\bb}_k\|^2)$ bits less than the naive approach in order to achieve the same distortion. This improvement can be made unbounded by choosing $\bB$ appropriately.
\end{example}

\begin{example}[Compressing observations of correlated relays]
\label{ex:relay}

Consider the problem of distributively compressing a $K$-dimensional Gaussian source $\bx$ with zero mean and covariance matrix $\bK_{\bx\bx}=\Tsnr\bH\bH^T+\bI$ for some $\Tsnr>0$ and some matrix $\bH\in\RR^{K\times K}$. This choice of covariance matrix corresponds to the joint distribution of the signals observed by $K$ relays in the Gaussian network depicted in Figure~\ref{fig:GaussianNetwork}, where it is assumed that each of the $K$ transmitters uses a random i.i.d. Gaussian codebook such that each of the signals $\bs_1,\ldots,\bs_K$ behaves statistically as white Gaussian noise. This network will be studied in more detail in the next subsection.

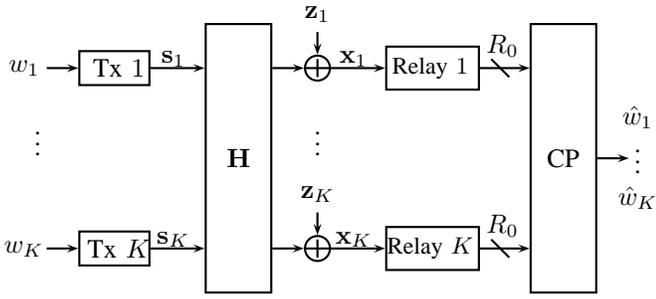
\begin{figure}[]
\begin{center}
\psset{unit=0.6mm}
\begin{pspicture}(0,0)(140,70)

\rput(0,50){
\rput(0,0){$w_1$} \psline{->}(5,0)(12,0)
\psframe(12,-4)(28,4)
\rput(21,0){$\text{Tx} \ 1$}
\psline{->}(28,0)(40,0)
\rput(33,2){$\bs_1$}
}

\rput(0,33){\rput(3,2){$\vdots$}}

\rput(0,10){\rput(0,0){$w_K$} \psline{->}(5,0)(12,0)
\psframe(12,-4)(28,4)
\rput(21,0){$\text{Tx} \ K$}
\psline{->}(28,0)(40,0)
\rput(33,2){$\bs_K$}
}

\rput(40,0){
\psframe(0,0)(15,60)\rput(7.5,30){$\bH$}}

\rput(55,0){
\rput(0,50){
\psline{->}(0,0)(7,0)
\rput(-18,0){
\pscircle(28,0){3}
\psline(25,0)(31,0)
\psline(28,-2)(28,2)
\rput(31,0){\psline{->}(0,0)(12,0)\rput(5,2){$\bx_1$}}
\psline{->}(28,8)(28,3)
\rput(28,12){$\bz_1$}
\psframe(43,-5)(64,5)
\rput(53,0){\small$\text{Relay} \ 1$}
\psline{->}(64,0)(75,0)
\psline(66.5,2)(70.5,-2)
\rput(69,5){$R_0$}
}
}

\rput(10,35){$\vdots$}

\rput(0,10){
\psline{->}(0,0)(7,0)
\rput(-18,0){
\pscircle(28,0){3}
\psline(25,0)(31,0)
\psline(28,-2)(28,2)
\rput(31,0){\psline{->}(0,0)(12,0)\rput(5,2){$\bx_K$}}
\psline{->}(28,8)(28,3)
\rput(28,12){$\bz_K$}
\psframe(43,-5)(64,5)
\rput(53,0){\small$\text{Relay} \ K$}
\psline{->}(64,0)(75,0)
\psline(66.5,2)(70.5,-2)
\rput(69,5){$R_0$}
}
}
}

\rput(112,0){
\psframe(0,0)(15,60)\rput(7.5,30){$\text{CP}$}
\psline{->}(15,30)(22,30)
\rput(25,30){$\begin{array}{c}
               \hat{w}_1 \\
               \vdots \\
               \hat{w}_K
             \end{array}$
}
}

\end{pspicture}
\end{center}
\caption{A Gaussian network with $K$ users and $K$ relays. Each relay sees one output of the channel $\bx=\bH\bs+\bz$ and has a clean bit-pipe of $R_0$ bits/channel use to the central processor (CP). The CP tries to estimate the messages transmitted by the $K$ users.}
\label{fig:GaussianNetwork}
\end{figure} 

\begin{figure*}[!ht]
\begin{center}
\subfloat[$\bH\in\RR^{4\times 4}$ with i.i.d. $\mathcal{N}(0,1)$ entries]{
\includegraphics[width=0.45\textwidth]{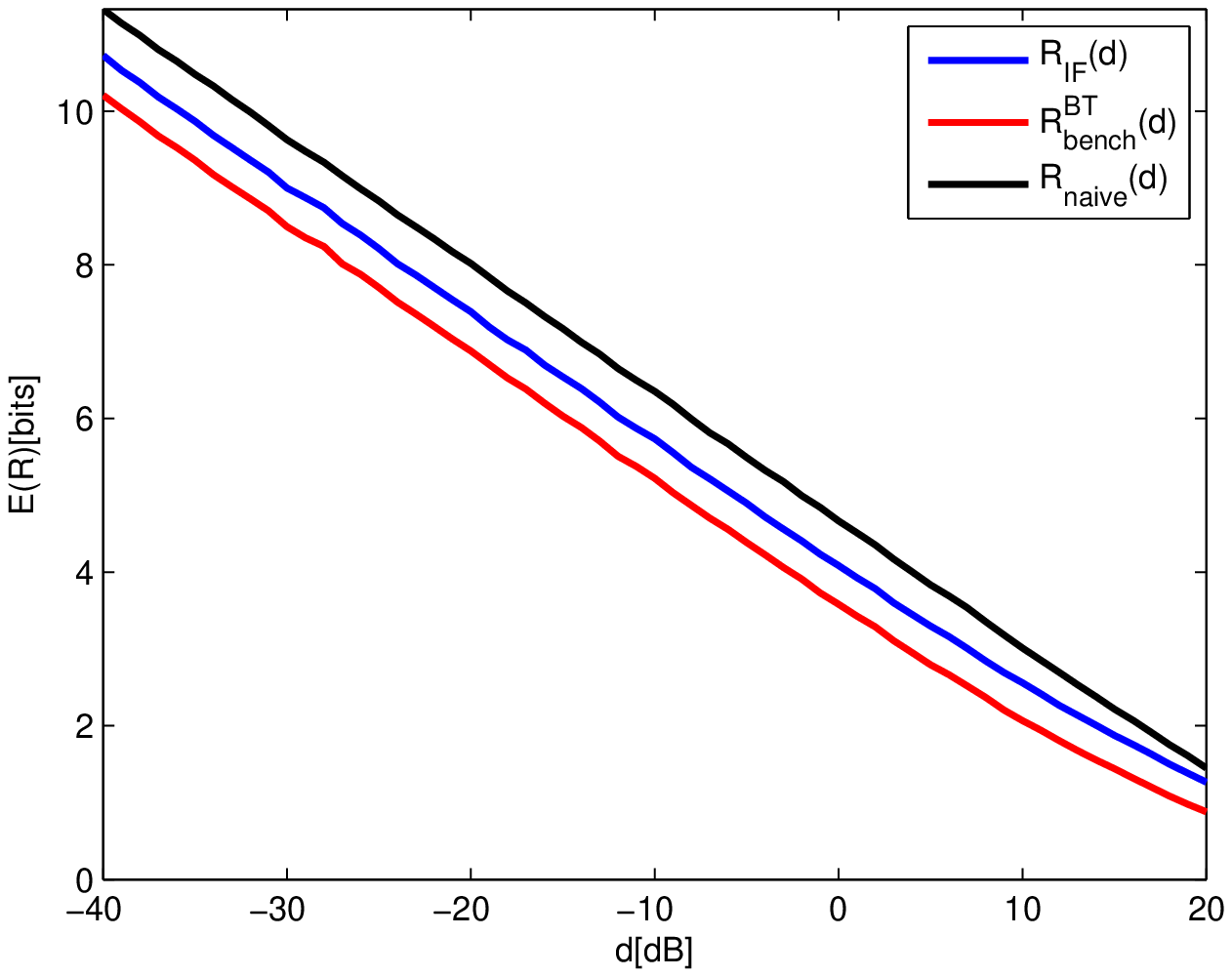}
\label{fig:RdK4M4SNR20dB}}
\qquad
\subfloat[$\bH\in\RR^{8\times 2}$ with i.i.d. $\mathcal{N}(0,1)$ entries]{
\includegraphics[width=0.45\textwidth]{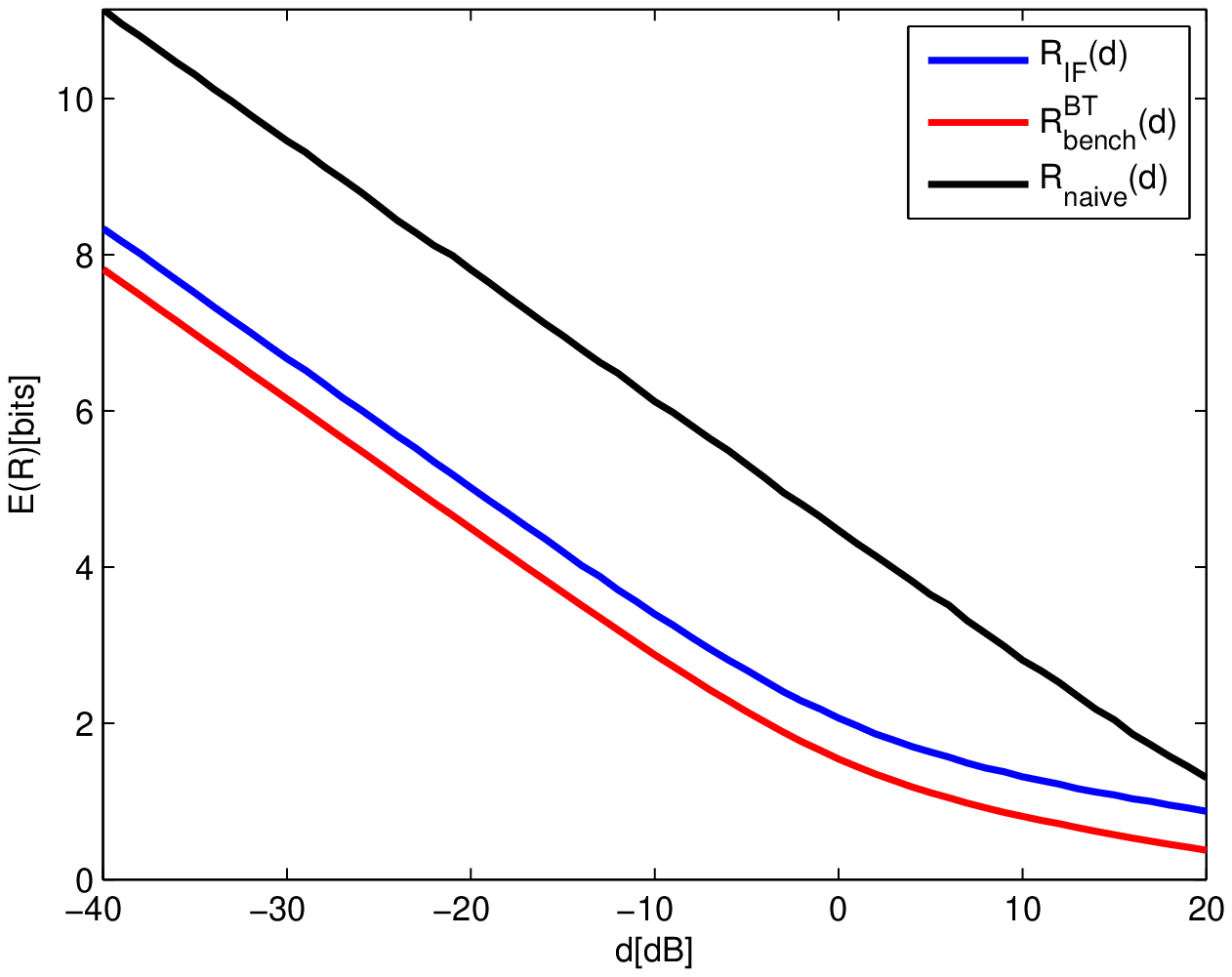}
\label{fig:RdK8M2SNR20dB}}

\end{center}
\caption{Comparison between the ergodic rates for the various compression schemes from Example~\ref{ex:relay}.}
\end{figure*}

We plot the averages of the minimal required compression rates for the two schemes, i.e. the ergodic rate-distortion functions of the two schemes, along with the ergodic benchmark rate-distortion function, under the assumption that the entries of $\bH$ are i.i.d. standard normal random variables. Figure~\ref{fig:RdK4M4SNR20dB} depicts these rates for $K=4$ and $\Tsnr=20$dB as a function of $d$. It is seen that at moderate to high resolution (small to moderate values of $d$) IF source coding closes about half of the gap between the naive compression scheme and the benchmark which corresponds to the Berger-Tung compression scheme.

One can argue that in the considered scenario the gap between the performance of the naive scheme and the benchmark is quite small, and therefore it is not clear if IF source coding only slightly improves over the naive scheme, or closely follows the performance of the Berger-Tung benchmark. To illustrate that the latter is true, in Figure~\ref{fig:RdK8M2SNR20dB} we consider a similar scenario where now $\bH\in\RR^{8\times 2}$ with i.i.d. $\mathcal{N}(0,1)$ entries. This models a network with $2$ transmitters and $8$ relays. This choice of distribution tends to induce more correlation between the entries of $\bx$, which enlarges the performance gap between Berger-Tung's compression and the naive compression approach. Nevertheless, as seen from Figure~\ref{fig:RdK8M2SNR20dB}, the gap between the performance of the Berger-Tung benchmark and IF source coding remains approximately the same.

\end{example}

\subsection{Layered Gaussian relay network}
\label{subsec:gaussnetwork}
In this subsection we consider the Gaussian network from Figure~\ref{fig:GaussianNetwork}, and show that for a wide regime of parameters using IF source coding as a building block improves upon other competing low-complexity coding schemes.

The Gaussian network we consider consists of $K$ non-cooperating transmitters, each with message $w_k$ and rate $R_k$. A central processor (CP) is interested in decoding all $K$ messages. However, it does not have a direct access to the signals transmitted by the $K$ transmitters. Instead, there are $K$ relays, each of which observes a noisy linear combination of the transmitted signals. Each relay has a clean bit-pipe of rate $R_0$ bits/channel use connecting it to the CP which it uses for helping the CP decode all messages.

Let $\bs_k\in\RR^{1\times n}$ be the signal transmitted by the $k$th transmitter during $n$ consecutive channel uses. We assume all transmitters are subject to the same power-constraint such that $\mathbb{E}\|\bs_k\|^2\leq n\Tsnr$ for all $k=1,\ldots,K$. Let $\bx_k\in\RR^{1\times n}$ be the signal received by the $k$th relay during $n$ consecutive channel uses, and let $\bS=[\bs_1^T \ \cdots \ \bs_K^T]^T$ and $\bX=[\bx_1^T \ \cdots \ \bx_K^T]^T$. The signals are related by
\begin{align}
\bX=\bH\bS+\bZ,\label{networkmodel}
\end{align}
where $\bH\in\RR^{K\times K}$ is the channel matrix between the $K$ transmitters and the $K$ relays and the entries of $\bZ\in\RR^{K\times n}$ are i.i.d. $\mathcal{N}(0,1)$. We are interested in the maximal achievable sum-rate $R_{\text{sum}}=\sum_{k=1}^K R_k$.

Clearly, $R_{\text{sum}}$ cannot exceed the MIMO capacity\footnote{Here, by capacity we mean the mutual information corresponding to a white input, as the transmitters are non-cooperating.} corresponding to the channel~\eqref{networkmodel} between the transmitters and relays, and it also cannot exceed $KR_0$ because even if each relay could decode all messages, the $K$ relays cannot convey more than $KR_0$ bits/channels use to the CP through the bit-pipes. Thus, we have
\begin{align}
R_{\text{sum}}\leq R_{\text{MIMO}}\triangleq\min\left(\frac{1}{2}\log|\bI+\Tsnr\bH\bH^T|,KR_0 \right).\label{mimoub}
\end{align}
An inner bound for $R_{\text{sum}}$ can be attained by the following scheme. Each relay can compress its observation $\bx_k$ with rate $R_0$ and send the compression index to the CP. The CP obtains $K$ estimates $\hat{\bx}_k=\bx_k+\bd_k$ of the relays' observations, where $\bd_k\in\RR^{1\times n}$ is the quantization error, and can use these estimates in order to decode the desired messages. Specifically, using this approach the CP decodes the messages from
\begin{align}
\hat{\bX}=\bH\bS+\bZ+\bD,\label{CPsignal}
\end{align}
where $\bD=[\bd_1^T \ \cdots \ \bd_K^T]^T$. If the quantization errors are statistically independent of everything else, as in IF source coding, $\bD$ can be treated as another additive noise. Let
\begin{align}
d(R_0)=\max_{k=1,\ldots,K}\frac{1}{n}\mathbb{E}(\|\bd_k\|^2).\nonumber
\end{align}
Assuming that all transmitters use i.i.d. Gaussian codebooks, it follows from the entropy power inequality~\cite[Problem 9.21]{coverthomas} that the CP can decode all messages $w_1,\ldots,w_K$ from the channel~\eqref{CPsignal} if
\begin{align}
R_{\text{sum}}\leq\frac{1}{2}\log\left|\bI+\frac{\Tsnr}{1+d(R_0)}\bH\bH^T\right|
\end{align}
Clearly, the degradation of this scheme w.r.t. the MIMO capacity depends on the value of $d(R_0)$. Improving the compression scheme decreases $d(R_0)$ which in turn increases $R_{\text{sum}}$. One can use the conditionally unbiased version of Berger-Tung in order to obtain a small $d(R_0)$. However, this solution requires joint typicality decoding at the CP which is difficult to implement. Alternatively, IF source coding can be employed, which considerably reduces the implementation complexity at the price of slightly increasing $d(R_0)$. The relays can also employ naive conditionally unbiased compression, which is also a low-complexity scheme. This reduces to performing IF source coding with the choice $\bA=\bI$ which is often suboptimal. The latter approach is often termed compress-and-forward in the literature~\cite{ng11IT}.

Alternatively, instead of compressing their noisy observations, the relays can attempt to decode the transmitted messages, or a function of the transmitted messages. In the decode-and-forward scheme~\cite{ce79} each relay decodes one of the messages and forwards this message to the CP. The compute-and-forward scheme~\cite{ng11IT} generalizes decode-and-forward and allows each relay to decode a linear combination of the messages, which is forwarded to the CP. Since decode-and-forward is a special case of compute-and-forward, its performance is never better.

\begin{figure*}[!t]
\begin{center}
\subfloat[$R_0=2$]{
\includegraphics[width=0.45\textwidth]{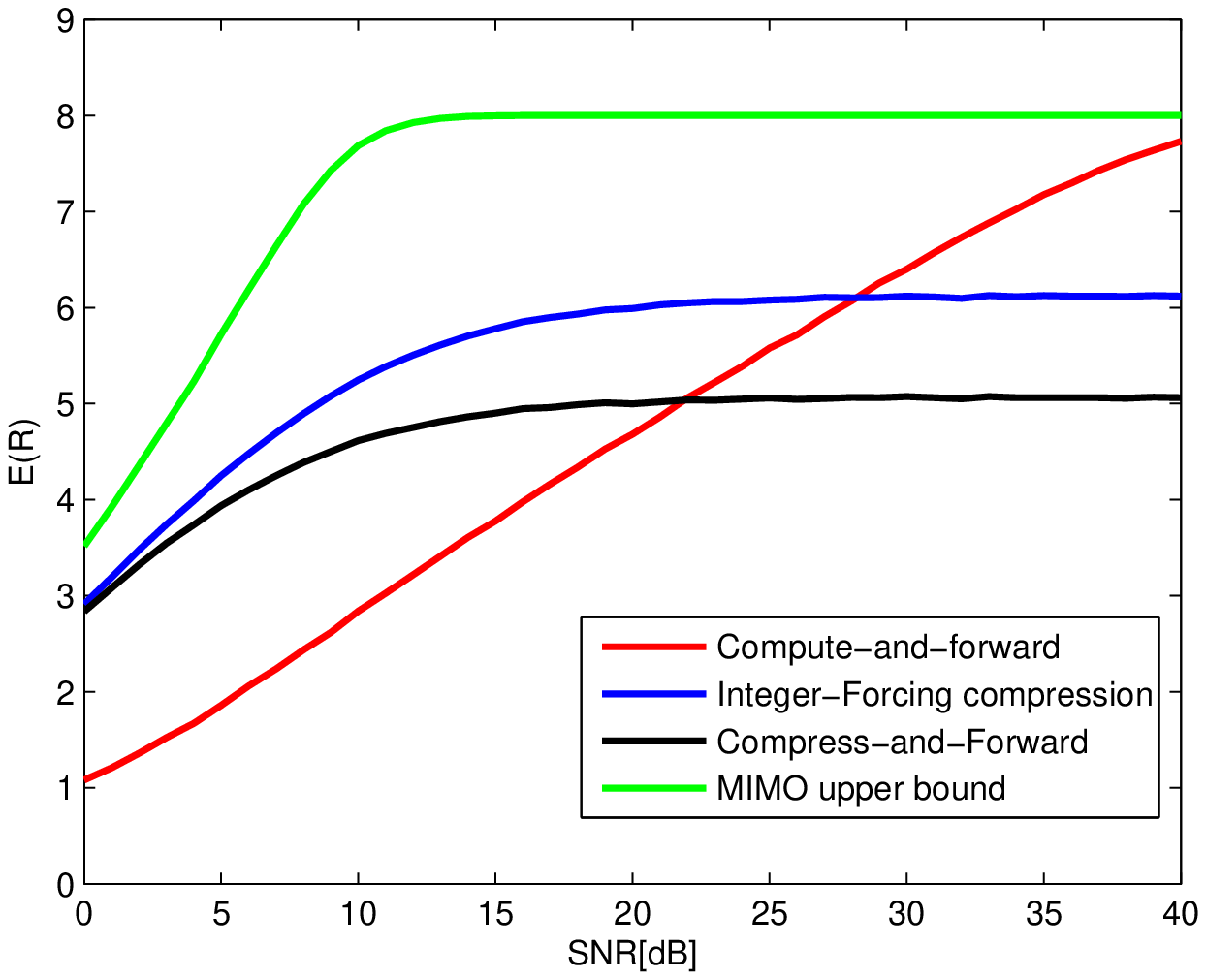}
\label{fig:networkK4R2}}
\qquad
\subfloat[$R_0=3$]{
\includegraphics[width=0.45\textwidth]{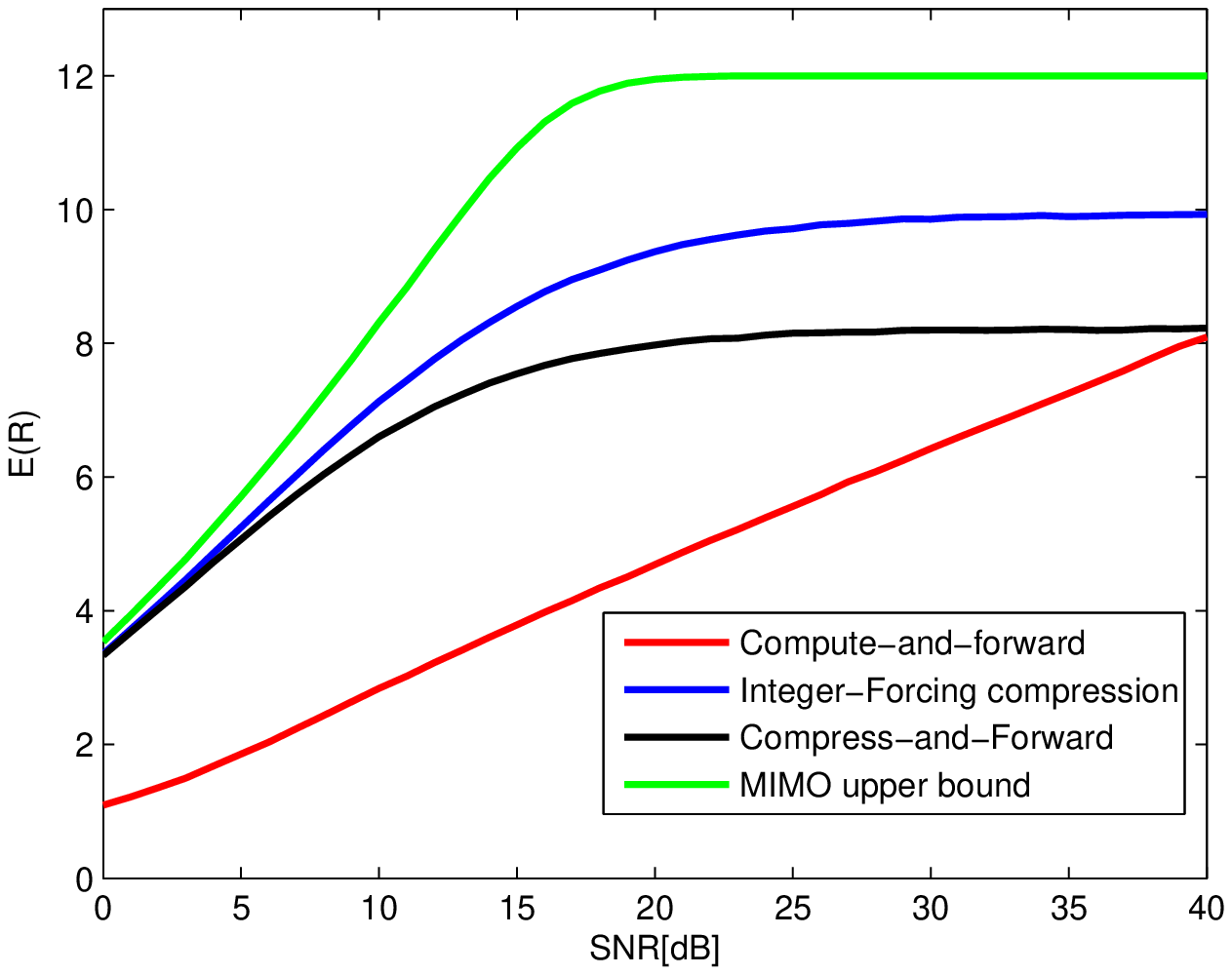}
\label{fig:networkK4R3}}

\end{center}
\caption{Ergodic rates over the network from Figure~\ref{fig:GaussianNetwork} for $K=4$}
\end{figure*}

In Figure~\ref{fig:networkK4R2} we plot the ergodic rates achieved using IF source coding, compress-and-forward and compute-and-forward, over the Gaussian network from Figure~\ref{fig:GaussianNetwork} For $R_0=2$ and $K=4$, where the entries of $\bH$ are assumed i.i.d. $\mathcal{N}(0,1)$. Figure~\ref{fig:networkK4R3} depicts the same ergodic rates for $R_0=3$.

The figures demonstrate that while compute-and-forward outperforms both compression-based schemes when $R_0$ is the system's bottleneck, for relatively large $R_0$ (w.r.t. the $1/K$ times the MIMO capacity) compression is preferable over decoding. The gains of IF source coding over naive compression are evident.

One can further improve performance using a quantize-map-and-forward like scheme~\cite{ssps09,adt11} where each relay quantizes its observation, bins it, and sends the bin index to the CP. The difference between such schemes and the compression based schemes described above is that in quantize-map-and-forward the CP decodes the messages from the bin indices themselves without ``decompressing'' the relays' observations. Such an approach improves upon compression based schemes. However, to date it lacks a signal processing based architecture allowing to reduce the problem to multiple instances of a point-to-point problem, as is the case for IF source coding. We note however that progress in the direction of developing a low-complexity architecture for quantize-map-and-forward has been made in~\cite{nwmt13}.

\subsection{Distributed joint source-channel coding}

\begin{figure}[]
\begin{center}
\psset{unit=0.6mm}
\begin{pspicture}(0,0)(92,70)

\rput(0,50){
\rput(0,0){$\bx_1$} \psline{->}(5,0)(12,0)
\psframe(12,-4)(28,4)
\rput(21,0){$\mathcal{E}_1$}
\psline{->}(28,0)(40,0)
\rput(33,3){$\bs_1$}
}

\rput(0,33){\rput(3,2){$\vdots$}}

\rput(0,10){\rput(0,0){$\bx_K$} \psline{->}(5,0)(12,0)
\psframe(12,-4)(28,4)
\rput(21,0){$\mathcal{E}_K$}
\psline{->}(28,0)(40,0)
\rput(33,3){$\bs_K$}
}

\rput(40,0){
\rput(0,50){
\pscircle(3,0){3}
\psline(0,0)(6,0)
\psline(3,-2)(3,2)
\psline{->}(3,8)(3,3)
\rput(3,12){$\bz_1$}
\psline{->}(6,0)(20,0)
\rput(12,3){$\by_1$}
}
\rput(3,35){$\vdots$}

\rput(0,10){
\pscircle(3,0){3}
\psline(0,0)(6,0)
\psline(3,-2)(3,2)
\psline{->}(3,8)(3,3)
\rput(3,12){$\bz_K$}
\psline{->}(6,0)(20,0)
\rput(12,3){$\by_K$}
}
}

\rput(60,0){
\psframe(0,0)(15,60)\rput(7.5,30){$\mathcal{D}$}
\psline{->}(15,30)(22,30)
\rput(25,30){$\begin{array}{c}
               \hat{\bx}_1 \\
               \vdots \\
               \hat{\bx}_K
             \end{array}$
}
}

\end{pspicture}
\end{center}
\caption{A distributed joint source-channel coding setting. Each encoder wishes to describe its observation $\bx_k$ to the decoder through an AWGN channel, with minimal average MSE distortion. The sources are correlated and the encoders are distributed.}
\label{fig:jscc}
\end{figure}
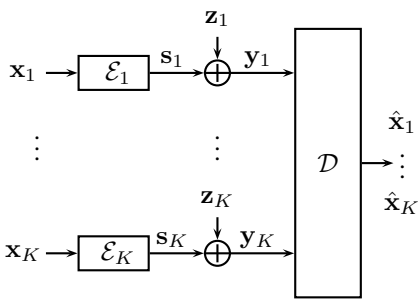 

In this subsection we consider the setup depicted in Figure~\ref{fig:jscc}. In this setup, there are $K$ distributed encoders, each with access to the vector $\bx_k$ that contains $n$ i.i.d. samples of the random variable $x_k$. We assume that the random vector $\bx=[x_1 \ \cdots \ x_K]^T$ is a Gaussian vector with zero mean and covariance matrix $\bK_{\bx\bx}$. Each encoder is equipped with an encoding function $\mathcal{E}_k:\RR^n\rightarrow\RR^n$, such that the signal it transmits to the decoder is $\bs_k=\mathcal{E}_k(\bx_k)$. All encoders are subject to the same power constraint $\mathbb{E}(\|\bs_k\|^2)=n P$. The decoder observes the transmitted signals through $K$ parallel AWGN channels
\begin{align}
\by_k=\bs_k+\bz_k, \ \ \ k=1,\ldots,K \nonumber
\end{align}
where the entries of $\bz_1,\ldots,\bz_K$ are i.i.d. Gaussian random variables with zero mean and variance $N$. The decoder has $K$ functions $\mathcal{D}_k:\RR^n\times\cdots\times\RR^n\rightarrow\RR^n$ that it uses in order to form estimates $\hat{\bx}_k=\mathcal{D}_k(\by_1,\ldots,\by_K)$ for each source.

Let $\Tsnr\triangleq P/N$. An $\Tsnr$-distortion vector $(\Tsnr,d_1,\ldots,d_K)$ is achievable if there exist encoding functions $\mathcal{E}_1,\ldots,\mathcal{E}_K$ and decoding functions $\mathcal{D}_1,\ldots,\mathcal{D}_K$ such that
\begin{align}
\frac{1}{n}\mathbb{E}\left(\|\bx_k-\hat{\bx}_k\|^2\right)\leq d_k,\label{distcond1}
\end{align}
for all $k=1,\ldots,K$. A conditionally \emph{unbiased} $\Tsnr$-distortion vector $(\Tsnr,d_1,\ldots,d_K)$ is achievable if in addition to~\eqref{distcond1}, the condition
\begin{align}
\mathbb{E}(\hat{\bx}_k|\bX)=\bx_k, \ \ k=1,\ldots,K\label{biascond1}
\end{align}
is satisfied. As before, we restrict attention to conditionally unbiased estimates, and focus on the maximal distortion among the $K$ vectors, i.e., $d=\max_{k=1,\ldots,K}d_k$.

An obvious approach for the considered problem is \emph{separation} of source coding and channel coding. This corresponds to using AWGN capacity achieving codebooks for transforming the $K$ AWGN channels into $K$ bit-pipes each with capacity $C=1/2\log(1+\Tsnr)$ bits/channel use, and then using distributed source coding with rate $C$ bits/sample at each encoder in order to describe the sources to the decoder. The main drawback of this approach is that it must be designed for specific values of $\Tsnr$ and required distortions $d_1,\ldots,d_K$. The predefined $\Tsnr$ acts as a threshold. If the actual $\Tsnr$ experienced by the communication system turns out to be higher than this threshold, the expected distortions would be $d_1,\ldots,d_k$, but would not improve when the actual $\Tsnr$ is improved.

Taking $K=1$ in our setup reduces it to a point-to-point problem of Gaussian source transmission over an AWGN channel. It is well known~\cite{Goblick65} that analog transmission of the source with appropriate scaling at the encoder and decoder achieves the optimal performance. Moreover, the transmitter's operation does not depend on the noise's variance at the receiver. As a result, if the noise variance turns out to be smaller than expected, the decoder can improve the quality of its estimate for the source. This desirable phenomena was extended to the Wyner-Ziv/dirty-paper setting in~\cite{kz09}. Here, we use the idea of IF source coding for constructing a joint source-channel coding scheme for our setup with an arbitrary number of users. The encoders' operation in the proposed scheme is independent of the noise variance, and the obtained expected distortion at the decoder decreases with $N$, provided that $N$ is below some predefined threshold.

The proposed coding approach utilizes a single lattice $\Lambda$ with $\sigma^2(\Lambda)=P$, that is good for channel coding and for MSE quantization. In particular, its goodness for MSE quantization implies that $r^2_{\text{eff}}(\Lambda)/n\approx P$. The coding scheme is designed assuming that the AWGN variance is not greater than some nominal value $N^{\text{nom}}$. However, when $N<N^{\text{nom}}$, the obtained distortion decreases as $N$ decreases.

Each encoder scales its observation by some $\beta>0$ to be defined shortly,\footnote{In general, performance can be improved by letting each encoder use a different $\beta_k$. We disregard this possibility for simplicity of exposition.} adds a dither $\bd_k$ uniformly distributed over $\CV$, and reduces the result \hspace{-4mm} $\mod\Lambda$ such that the transmitted signals are
\begin{align}
\bs_k=\left[\beta\bx_k+\bd_k\right]\Mod, \ \ \ k=1,\ldots,K.\nonumber
\end{align}
Note that the power constraints are satisfied as $\bs_k$ is uniformly distributed over $\CV$ and therefore its second moment equals $\sigma^2(\Lambda)$. The decoder first performs MMSE estimation of each $\bs_k$, by scaling each $\by_k$ by $\alpha=\sqrt{P/(P+N)}$, subtracting back the dither and reducing  \hspace{-3mm} $\mod\Lambda$. This gives
\begin{align}
\tilde{\by}_k&=\left[\alpha\by_k-\bd_k\right]\Mod\nonumber\\
&=\left[\bs_k+(\alpha-1)\bs_k+\alpha\bz_k-\bd_k\right]\Mod\nonumber\\
&=\left[\beta\bx_k+\bz_{\text{eff},k}\right]\Mod,\nonumber
\end{align}
where
\begin{align}
\bz_{\text{eff},k}\triangleq (\alpha-1)\bs_k+\alpha\bz_k.\nonumber
\end{align}
The noise $\bz_{\text{eff},k}$ is statistically independent of $\bx_k$, and has effective variance of
\begin{align}
\frac{1}{n}\mathbb{E}(\|\bz_{\text{eff},k}\|^2)=\frac{NP}{N+P}, \ \ k=1,\ldots,K.\nonumber
\end{align}
Moreover, it is a linear combination of a dither uniformly distributed over the Voronoi region of a lattice that is good for MSE quantization and an AWGN, and therefore, by Lemma~\ref{lem:mixturenoise}, it is semi-norm ergodic.

As before, let $\bX=[\bx_1^T \ \cdots \ \bx_K^T]^T$, and define $\tilde{\bY}$ and $\bZ_{\text{eff}}$ in a similar manner. The decoder chooses a full-rank matrix $\bA\in\ZZ^{K\times K}$ and computes
\begin{align}
\widehat{\beta\bA\bX}&\triangleq\left[\bA\tilde{\bY} \right]\Mod\nonumber\\
&=\left[\bA(\left[\beta\bX+\bZ_{\text{eff}}\right]\Mod) \right]\Mod\nonumber\\
&=\left[\bA(\beta\bX+\bZ_{\text{eff}}) \right]\Mod.\nonumber
\end{align}
Let $\ba_k^T$ be the $k$th row of $\bA$. The random vector $\ba_k^T(\beta\bX+\bZ_{\text{eff}})$ is semi-norm ergodic with zero mean and effective variance
\begin{align}
\sigma^2_k&\triangleq\frac{1}{n}\mathbb{E}(\|\ba_k^T(\beta\bX+\bZ_{\text{eff}})\|^2)\nonumber\\
&=\ba_k^T\left(\beta^2\bK_{\bx\bx}+\frac{NP}{N+P}\bI\right)\ba_k.\nonumber
\end{align}
Since $\Lambda$ is good for channel coding, if $\sigma^2_k<P$ for all $k=1,\ldots,K$, then
\begin{align}
\widehat{\beta\bA\bX}\stackrel{(w.h.p.)}{=}\bA(\beta\bX+\bZ_{\text{eff}}),\label{whpeq2}
\end{align}
and the decoder can further compute
\begin{align}
\hat{\bX}&=\frac{1}{\beta}A^{-1}\widehat{\beta\bA\bX}\nonumber\\
&\hspace{-4mm}\stackrel{(w.h.p.)}{=}\bX+\frac{1}{\beta}\bZ_{\text{eff}},\nonumber
\end{align}
which are unbiased estimates of each $\bx_k$ with average MSE distortion of $d_{\text{IF}}=NP/\beta^2(N+P)$.

The remaining question is how to choose $\beta$ such that~\eqref{whpeq2} indeed holds. Recall that $\beta$ is chosen by the encoders that only know that $N<N^{\text{nom}}$, rather than the exact value of $N$. Therefore, the encoders should choose $\beta$ as
\begin{align}
&\beta_{\text{opt}}(P,N^{\text{nom}},\bK_{\bx\bx})\triangleq\nonumber\\
& \max_{\beta>0} s.t. \min_{\substack{{\bA\in\ZZ^{K\times K}}\\ {\det(\bA)\neq 0}}}\max_{k=1,\ldots,K}\ba_k^T(\beta^2\bK_{\bx\bx}+\frac{N^{\text{nom}}P}{N^{\text{nom}}+P}\bI)\ba_k =P\nonumber
\end{align}
and the symmetric distortion obtained by the proposed scheme is
\begin{align}
d_{\text{IF}}=\frac{N}{\beta^2_{\text{opt}}(P,N^{\text{nom}},\bK_{\bx\bx})}\nonumber
\end{align}
which decreases as $N$ decreases, as desired.

A naive joint source-channel coding schemes that ignores the correlations between the entries of $\bx$ would be transmitting each $x_k$ in an analog Goblick-like scheme. The distortion achieved by such a scheme would be\footnote{Taking into account the constraint that the estimate for each $x_k$ must be conditionally unbiased.}
\begin{align}
d_{\text{naive}}=\frac{N}{P}\max_{k=1,\ldots,K}\bK_{\bx\bx}(k,k).\nonumber
\end{align}
It can be easily verified that the same distortion is achieved if one constrains $\bA=\bI$ in the scheme proposed here. Therefore, the proposed IF based joint source-channel coding scheme strictly improves upon the naive one.

It is also worth mentioning that the proposed scheme easily generalizes to a dirty paper scenario, where the output of each AWGN channel is further corrupted by an arbitrary interference $\bv_k$ known to encoder $k$ but not to the decoder, i.e., $\by_k=\bs_k+\bv_k+\bz_k$. In the proposed scheme, the encoders can transmit $\bs_k=[\beta\bx_k-\bv_k+\bd_k]\Mod$ and the decoder remains the same.

\section{One-shot Integer-Forcing Source Coding}
\label{sec:oneshot}
One of the advantages of IF source coding is that its complexity and performance can be traded-off, by choosing nested lattice codes that can be easily implemented, but are less effective as channel codes and MSE quantizers.

In the previous sections we have considered the extreme case of high-dimensional pairs of nested lattices where the fine lattice is good for MSE quantization and the coarse lattice is good for channel coding. In this section we consider the other extreme, where both lattices are scaled versions of the integer lattice $\ZZ$. With this choice of nested lattice pair, IF source coding becomes extremely easy to implement. Moreover, this one-shot version of IF source coding does not induce any latency and does not assume the existence of an unlimited number of i.i.d. samples to be compressed.

Let $\Lambda_f=\sqrt{12d}\ZZ$ and $\Lambda=2^R\sqrt{12d}\ZZ$. If $2^{R}$ is a positive integer then $\Lambda\subseteq\Lambda_f$, and the codebook $\mathcal{C}=\Lambda_f\cap\CV$ with rate $R$ is a valid codebook for IF source coding. Let $d_k$ be a random dither uniformly distributed over $\CV_f$, known to both the $k$th encoder and the decoder. The $k$th encoder conveys the index corresponding to the point
\begin{align}
\left[Q_{\Lambda_f}(x_k+d_k)\right]\Mod\nonumber
\end{align}
to the decoder. Note that for a 1D lattice, the quantization operation reduces to a simple slicer. Thus all operations are easy to implement.

The decoder first subtracts back the dither and reduces $\mod\hspace{0.1mm}\Lambda$ to obtain
\begin{align}
\tilde{x}_k\stackrel{(i.d.)}{=}\left[x_k+d_k\right]\Mod,\nonumber
\end{align}
and then chooses some full-rank matrix $\bA\in\ZZ^{K\times K}$ and computes
\begin{align}
\widehat{\bA\bx}\triangleq\left[\bA\tilde{\bx}\right]\Mod=\left[\bA(\bx+\bd)\right]\Mod,\label{uncodedest}
\end{align}
where $\bd=[d_1 \ \cdots \ d_K]^T$. In contrast to the case of a high-dimensional nested lattice codebook, where the probability that $\widehat{\bA\bx}\neq\bA\bx$ could be made as low as desired if $r_{\text{eff}}^2(\Lambda)$ is large enough, here this probability is finite for any finite value of $2^R\sqrt{12d}$. In particular, let $\ba_k^T$ be the $k$th row of $\bA$ and define the random variable
\begin{align}
w_k\triangleq\ba_k^T(\bx+\bd)\nonumber
\end{align}
with zero mean and variance
\begin{align}
\sigma^2_{w,k}=\ba_k^T(\bK_{\bx\bx}+d\bI)\ba_k.\nonumber
\end{align}
We have
\begin{align}
\Pr\left(\widehat{\bA\bx}\neq\bA\bx \right)&=\Pr\left(\bigcup_{k=1}^K [w_k]\Mod\neq w_k \right)\nonumber\\
&=\Pr\left(\bigcup_{k=1}^K Q_{\Lambda}(w_k)\neq 0 \right)\nonumber\\
&=\Pr\left(\bigcup_{k=1}^K |w_k|\geq \frac{1}{2}2^R\sqrt{12d} \right)\nonumber\\
&\leq\sum_{k=1}^K\Pr\left( |w_k|\geq 2^R\sqrt{3d} \right),\label{PeTmp}
\end{align}
where the last inequality follows from the union bound. Next, we apply the following Lemma from~\cite{fsk11,oe13}

\vspace{2mm}

\begin{lemma}{\cite[Lemma 3]{oe13}}
\label{lem:mixture}
Consider the random variable
\begin{align}
z_{\text{eff}}=\sum_{\ell=1}^{L} \alpha_\ell z_\ell +\sum_{k=1}^K \beta_k d_k\nonumber
\end{align}
where $\left\{z_\ell\right\}_{\ell=1}^L$ are i.i.d. Gaussian random variables with zero mean and some variance $\sigma^2_z$ and $\left\{d_k\right\}_{k=1}^K$ are i.i.d. random variables, statistically independent of $\left\{z_\ell\right\}_{\ell=1}^L$, uniformly distributed over the interval $[-\rho/2,\rho/2)$ for some $\rho>0$. Let $\sigma^2_{\text{eff}}\triangleq\mathbb{E}(z^2_{\text{eff}})$. Then
\begin{align}
\Pr(z_{\text{eff}}>\tau)=\Pr(z_{\text{eff}}<-\tau)\leq\exp\left\{-\frac{\tau^2}{2\sigma^2_{\text{eff}}}\right\}.\nonumber
\end{align}
\end{lemma}

\vspace{2mm}

One can easily verify that $w_k$ satisfies the conditions of Lemma~\ref{lem:mixture} as $\ba_k^T\bx$ is a Gaussian random variable statistically independent of the dither vector $\bd$. Therefore, we can further bound~\eqref{PeTmp} as
\begin{align}
\Pr&\left(\widehat{\bA\bx}\neq\bA\bx \right)\leq\sum_{k=1}^K 2\exp\left\{-\frac{2^{2R}3d}{2\ba_k^T\left(\bK_{\bx\bx}+d\bI\right)\ba_k} \right\}\nonumber\\
&\leq 2K\exp\left\{-\frac{3}{2}2^{2\left(R-\frac{1}{2}\log(\max_{k=1,\ldots,K}\ba_k^T\left(\bI+\frac{1}{d}\bK_{\bx\bx}\right)\ba_k)\right)} \right\}\nonumber\\
&=2K\exp\left\{-\frac{3}{2}2^{2\left(R-R_{\text{IF}}(\bA,d)\right)}\right\},\label{poverload}
\end{align}
where $R_{\text{IF}}(\bA,d)$ is the minimum required rate for a IF source coding when a good nested lattice pair is used, as defined in Theorem~\ref{thm:IFrd}.
The decoder proceeds by computing
\begin{align}
\hat{\bx}=\bA^{-1}\widehat{\bA\bx}=\bx+\bd+\bA^{-1}\left(\widehat{\bA\bx}-\bA\bx \right).\label{xestoneshot}
\end{align}
Since $d_k$ is statistically independent of $\bx$ and $\mathbb{E}(d_k^2)=d$ for all $k=1,\ldots,K$, we see that provided that $\widehat{\bA\bx}=\bA\bx$ the one-shot version of IF source coding produces conditionally unbiased estimates of $x_k$ with distortion $d$. The probability that $\widehat{\bA\bx}=\bA\bx$ can be controlled by increasing $R-R_{\text{IF}}(\bA,d)$ which is the coding overhead w.r.t. to IF source coding with an optimal nested lattice pair. For instance, if $K=4$, taking $R=R_{\text{IF}}(\bA,d)+2$ results in $\Pr\left(\widehat{\bA\bx}\neq\bA\bx \right)\leq 3\cdot 10^{-10}$. The next theorem summarizes the discussion above.

\vspace{1mm}

\begin{theorem}[One-shot IF source coding]
\label{thm:oneshot}
Let $R_{\text{IF}}(d)$ be as defined in Theorem~\ref{thm:IFrd} and set $R=R_{\text{IF}}(d)+\Delta$ for some $\Delta>0$. If $2^R$ is a positive integer, the one-shot version of IF source coding with lattices $\Lambda_f=\sqrt{12d}\ZZ$ and $\Lambda=2^R\sqrt{12d}\ZZ$ produces conditionally unbiased estimates with average MSE distortion $d$ for each $x_k$, $k=1,\ldots,K$ with probability greater than $1-2K\exp\{-\frac{3}{2}2^{2\Delta}\}$.
\end{theorem}

\subsection{Modulo Analog-to-Digital Converters}
\label{subsec:modADCs}
Theorem~\ref{thm:oneshot} shows that a simple implementation of IF source coding with $1D$ lattices only requires a small rate overhead w.r.t. to the asymptotic performance of IF source coding. The simplicity of the one-shot IF source coding scheme suggests that this framework may be useful for designing Analog-to-Digital converters (ADCs) that can exploit correlations in a distributed manner. To illustrate the problem, consider the Gaussian MIMO channel $\bx=\bH\bs+\bz$, where $\bH\in\RR^{K\times M}$ is the channel matrix, $\bz\in\RR^{K\times 1}$ is a vector of AWGN and $\bs$ are the $M$ inputs to channel, which are assumed to be i.i.d. normally distributed. The front-end of the MIMO receiver consists of $K$ ADCs, one for the output of each receive antenna. Today, each of these ADCs is designed w.r.t. the \emph{marginal} distribution of each output, ignoring the fact that the $K$ ADCs sample correlated signals. Often, the variance of each output is quite large although the \emph{conditional} variance when all other samples are given is small. Thus, exploiting the spatial correlation may significantly reduce the distortion created by the ADCs. However, the ADCs are expected to work at very high rates, which precludes cooperation between their operations. We show that a variant of the one-shot IF source coding scheme allows the ADCs to exploit the spatial correlations with no cooperation and with roughly the same encoding complexity as a standard ADC, and only a small increase in the decoding complexity.

The one-shot version of IF source coding described above requires each encoder to first quantize its observation using a scaled integer lattice, and then reduce the result modulo the coarse lattice, which is also a scaled version of $\ZZ$. This can be implemented by applying an ADC as the quantizer followed by a digital modulo reduction. However, the power consumption and the complexity of an ADC are dictated by the number of bits it produces. Therefore, if the modulo operation can be implemented efficiently in the analog domain, performance can be improved by first applying the modulo reduction, and only then incorporating the ADC. Since the modulo reduced signal is of a smaller support, less bits are required for describing it with the same average distortion level. The next lemma shows that if $\Lambda_f=\sqrt{12d}\ZZ$ and $\Lambda=2^R\sqrt{12d}\ZZ$ the operations $\Qlf$ and $\Mod$ commute, i.e., one can first reduce the signal $\Mod$ and then quantize to $\Lambda_f$, rather than first quantizing and then reducing $\Mod$.

\vspace{1mm}

\begin{lemma}
\label{lem:commute}
Let $2^R$ be a positive odd integer and define the nested lattices $\Lambda=\sqrt{12d}\ZZ$ and $\Lambda_f=2^R\sqrt{12d}\ZZ$ for some $d>0$. for any $x\in\RR$ we have
\begin{align}
\left[\Qlf(x) \right]\Mod=\Qlf\left([x]\Mod \right).\nonumber
\end{align}
\end{lemma}

\vspace{1mm}

\begin{proof}
See Appendix~\ref{app:commuteproof}
\end{proof}

\vspace{1mm}

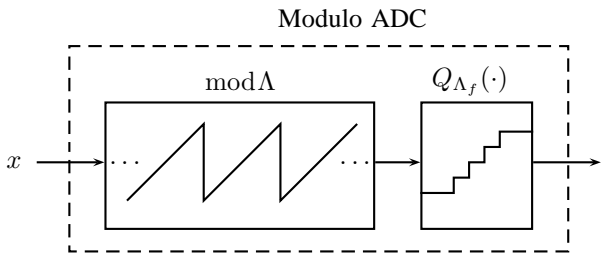
\begin{figure}[]
\begin{center}
\psset{unit=0.6mm}
\begin{pspicture}(0,10)(130,70)

\rput(-5,30){$x$} \psline{->}(0,30)(15,30)
\psframe(15,15)(75,43.5)
\rput(45,48){$\Mod$}
\psline(20,21.5)(37,38.5)(37,21.5)(54,38.5)(54,21.5)(71,38.5)
\rput(70.5,30){$\ldots$}
\rput(19.5,30){$\ldots$}

\rput(75,0){
\psline{->}(0,30)(10,30)
\psframe(10,15)(35,43.5)
\rput(21,48){$\Qlf(\cdot)$}
\rput(10,30){
\psline(0,-6.8)(7.4,-6.8)(7.4,-3.4)(10.8,-3.4)(10.8,0)(14.2,0)(14.2,3.4)(17.6,3.4)(17.6,6.8)(25,6.8)
}
}

\psline{->}(110,30)(125,30)

\rput(70,62){Modulo ADC}
\psframe[linestyle=dashed](7,10)(118,56)

\end{pspicture}
\end{center}
\caption{A schematic illustration of the modulo ADC for $2^R=5$. This component can act as an encoder in the one-shot version of IF source coding.}
\label{fig:analogADC}
\end{figure}

Lemma~\ref{lem:commute} implies that the $1D$ version of IF source coding can indeed be implemented by first reducing the source $x$ modulo $\Lambda$ and only then quantizing it to $\Lambda_f$. The advantage in switching the order of the operations is that if the $1D$ modulo reduction, which is equivalent to the ``saw-tooth'' function, can be efficiently implemented in the analog domain, then the quantizer that follows it can be implemented using an ADC with only $R$ bits/sample. The relation between $R$, the obtained distortion, and the error probability is characterized in Theorem~\ref{thm:oneshot} and depends on $R_{\text{IF}}(d)$. Figure~\ref{fig:analogADC} depicts the architecture of the proposed modulo ADC, that can replace the encoders in the one-shot IF source coding scheme.

\section{Summary and Conclusions}
\label{sec:sum}
We have presented and analyzed a new low-complexity framework for distributed lossy compression, which is based on the integer-forcing architecture. This framework allows the system designer to trade performance and complexity by appropriately choosing the nested lattice codebooks that are used. A remarkable feature of the proposed scheme is that it admits a very simple one-shot version, whose performance is not very far from that obtained using IF source coding with asymptotically good nested lattice codes. We have also shown that if one can implement the $1D$ modulo operation with an analog circuit, which corresponds to implementing the ``saw-tooth'' function, then the IF source coding approach can translate to a novel ADC design, suitable for sampling spatially correlated sources. Such ADCs can potentially be very useful for the front-end of a MIMO receiver, where standard ADC designs are already challenged by the growing transmission rates.

We remark that the IF equalization framework for Gaussian MIMO channels~\cite{zneg12IT} has been extended to an equalization framework for Gaussian intersymbol-interference channels~\cite{oe12IT}. In a similar manner, the IF source coding framework proposed here, which is suitable for distributed lossy compression of spatially correlated signals, can be extended to an IF compression framework for stationary temporally correlated signals. Nevertheless, such a solution is less attractive as one can always use a sequential Wyner-Ziv like compression scheme for a stationary source. In such a scheme the first samples of the source are compressed without binning/modulo reduction, and the next samples are first binned/modulo reduced and then compressed. The decoder uses the samples that are not binned for recovering the next samples in a sequential manner. This Wyner-Ziv scheme suffers from the intrinsic overhead of having to describe the first samples to the decoder without binning. This overhead can be made negligible by increasing the length of the compression block. For spatially correlated sources a similar Wyner-Ziv like compression scheme will result in asymmetric compression rates, which is a consequence of the lack of ``spatial stationarity''.

\begin{appendices}

\section{Proof of Lemma~\ref{lem:commute}}
\label{app:commuteproof}

We begin with two general Lemmas from which Lemma~\ref{lem:commute} is immediately deduced.

\begin{lemma}
\label{lem:commutecond}
For any pair of $n$-dimensional nested lattices $\Lambda\subseteq\Lambda_f$ and any $\bx\in\RR^n$
\begin{align}
\left[\Qlf(\bx)\right]\Mod&=\Qlf\left([\bx]\Mod\right)\nonumber\\
&+\Ql\left(\left[\Qlf(\bx)\right]\Mod+\bx-\Qlf(\bx) \right).\nonumber
\end{align}
\end{lemma}

\vspace{1mm}

\begin{proof}
\begin{align}
&\left[\Qlf(\bx)\right]\Mod=\Qlf(\bx)-\Ql\left(\Qlf(\bx)\right)\nonumber\\
&\ \ \ \ \ =\Qlf\left(\bx-\Ql(\bx)+\Ql(\bx)\right)-\Ql\left(\Qlf(\bx)\right)\nonumber\\
&\ \ \ \ \ =\Qlf\left(\bx-\Ql(\bx)\right)+\Qlf\left(\Ql(\bx)\right)-\Ql\left(\Qlf(\bx)\right)\nonumber\\
&\ \ \ \ \ =\Qlf\left([\bx]\Mod\right)+\Ql(\bx)-\Ql\left(\Qlf(\bx)\right),\label{tmpeq1}
\end{align}
where in the last equality we have used the fact that $\Qlf\left(\Ql(\bx)\right)=\Ql(\bx)$ since $\Lambda\subseteq\Lambda_f$.
We have,
\begin{align}
&\Ql(\bx)=\Ql\left(\Qlf(\bx)+\bx-\Qlf(\bx) \right)\nonumber\\
&=\Ql\left(\Qlf(\bx)-\Ql\left(\Qlf(\bx)\right)+\Ql\left(\Qlf(\bx)\right)+\bx-\Qlf(\bx) \right)\nonumber\\
&=\Ql\left(\left[\Qlf(\bx)\right]\Mod+\bx-\Qlf(\bx) \right)+\Ql\left(\Qlf(\bx)\right).\label{tmpeq2}
\end{align}
Substituting~\eqref{tmpeq2} in~\eqref{tmpeq1} gives the desired result.
\end{proof}

\vspace{1mm}

\begin{lemma}
\label{lem:tiling}
If the pair of nested lattices $\Lambda\subseteq\Lambda_f$ satisfies the tiling condition $\CV=\left(\Lambda_f\cap\CV\right)+\CV_f$ then
\begin{align}
\left[\Qlf(\bx)\right]\Mod&=\Qlf\left([\bx]\Mod\right)\nonumber.
\end{align}
for any $\bx\in\RR^n$.
\end{lemma}

\vspace{1mm}

\begin{proof}
For any $\bx\in\RR^n$ we have
\begin{align}
\left[\Qlf(\bx)\right]\Mod\in\left(\Lambda_f\cap\CV\right), \ \text{and} \ \bx-\Qlf(\bx)\in\CV_f.\nonumber
\end{align}
Therefore
\begin{align}
\left[\Qlf(\bx)\right]\Mod+\bx-\Qlf(\bx)\in\left(\Lambda_f\cap\CV\right)+\CV_f,\nonumber
\end{align}
The tiling condition $\CV=\left(\Lambda_f\cap\CV\right)+\CV_f$ implies that
\begin{align}
\left[\Qlf(\bx)\right]\Mod+\bx-\Qlf(\bx)\in\CV,\nonumber
\end{align}
which implies that
\begin{align}
\Ql\left(\left[\Qlf(\bx)\right]\Mod+\bx-\Qlf(\bx) \right)=0.\nonumber
\end{align}
The result now follows immediately from Lemma~\ref{lem:commutecond}.
\end{proof}

\vspace{1mm}

It is easy to verify that if $2^R$ is a positive odd integer the nested lattices $\Lambda=\sqrt{12d}\ZZ$ and $\Lambda_f=2^R\sqrt{12d}\ZZ$ satisfy the tiling condition $\CV=\left(\Lambda_f\cap\CV\right)+\CV_f$, and Lemma~\ref{lem:commute} immediately follows from Lemma~\ref{lem:tiling}

\end{appendices}

\bibliographystyle{IEEEtran}
\bibliography{IF_Source_Coding_bib}

\end{document}